\documentclass[aps,physrev,showpacs,twocolumn,amssymb,superscriptaddress,notitlepage,floats,floatfix,nofootinbib]{revtex4-2}
\pdfoutput=1
\usepackage{booktabs}
\usepackage{graphicx}
\usepackage{subfig}
\usepackage{amsmath}
\usepackage{amsfonts}
\usepackage{tensor}
\usepackage{bm}
\usepackage{mathrsfs}
\usepackage{orcidlink}
\usepackage{booktabs}

\usepackage{url}
\usepackage[utf8]{inputenc}

\usepackage{threeparttablex}

\usepackage{color}
\usepackage{multirow}
\usepackage{palatino}
\usepackage{float}
\usepackage{xcolor}

\usepackage{soul}
\usepackage{calrsfs}
\usepackage{fancyhdr}

\def\beq{\begin{equation}}
\def\eeq{\end{equation}}
\def\bear{\begin{eqnarray}}
\def\ear{\end{eqnarray}}

\def\L{\mathscr{L}}

\begin{document}

\title{Electrically charged regular black holes in higher dimensions:\\ Maxwellian limit, branch point of the Lagrangian, and thermodynamics}

\author{Bobir Toshmatov\,\orcidlink{0000-0002-7853-186X}}
\email{toshmatov@astrin.uz} 
\affiliation{New Uzbekistan University, Movarounnahr street 1, Tashkent 100000, Uzbekistan}
\affiliation{Ulugh Beg Astronomical Institute, Astronomy street 33, Tashkent 100052, Uzbekistan}

\author{Bobomurat Ahmedov\,\orcidlink{0000-0002-1232-610X}}
\email{ahmedov@astrin.uz}
\affiliation{Institute of Theoretical Physics, National University of Uzbekistan, Tashkent 100174, Uzbekistan}
\affiliation{School of Physics, Harbin Institute of Technology, Harbin 150001, China}

\author{Nozima Isamadinova}
\email{n.isamadinova@newuu.uz} 
\affiliation{Ulugh Beg Astronomical Institute, Astronomy street 33, Tashkent 100052, Uzbekistan}

\author{Ozodbek Rahimov}
\email{rahimov@astrin.uz} 
\affiliation{Tashkent State University of Economics, Islom Karimov street 49, 100066, Uzbekistan}
\affiliation{Ulugh Beg Astronomical Institute, Astronomy street 33, Tashkent 100052, Uzbekistan}

\author{Chengxun Yuan\,\orcidlink{0000-0002-2308-6703}}
\email{yuancx@hit.edu.cn}
\affiliation{School of Physics, Harbin Institute of Technology, Harbin 150001, China}

\begin{abstract}
Electrically charged regular black holes in general relativity coupled to nonlinear electrodynamics are known to require a double-valued Lagrangian density $\L(F)$, with a branch point at the radius where the electromagnetic invariant attains its extremum. We extend the construction of such black holes to an arbitrary number of spacetime dimensions $D\geq4$ and obtain a new family of solutions that unifies the Bardeen and Hayward black holes in a single analytic form, with a single nonlinearity parameter $\gamma$. We show that the Maxwellian weak-field limit is recovered if and only if $\gamma=D-3$, so that the Bardeen and Hayward black holes, which lack this limit in four dimensions, become Maxwellian in five and six dimensions, respectively. The matter sector is written in the dual $P$ framework as a Hamiltonian density $\mathcal H(P)$ that contains only fixed coupling constants, so that the mass and the charge of the black hole appear as integration constants of a fixed theory; the regular solutions form a one-parameter subfamily in which the mass is tied to the charge. We locate the branch point of $\L(F)$ in closed form and find that it is always hidden inside the event horizon in four dimensions, whereas for $D\geq5$ it emerges from the horizon for near-extremal charges and becomes visible to a distant observer. The null and weak energy conditions are satisfied everywhere, the strong energy condition is violated only inside the horizon, and the dominant energy condition is violated at large distances only for $\gamma>D-3$. Finally, since the theory is fixed, the first law of black hole thermodynamics holds in its standard form; we derive the horizon potential in closed form, the extended first law with the couplings as thermodynamic variables, and the corresponding Smarr formula.
\end{abstract}

\maketitle

\section{Introduction}

The singularity problem remains one of the most formidable challenges in theoretical physics. According to the singularity theorems of Penrose \cite{PhysRevLett.14.57} and Hawking \cite{Hawking:1973uf}, the gravitational collapse of physically reasonable matter in general relativity inevitably leads to a spacetime singularity, where the curvature becomes infinite and the predictive power of classical physics breaks down. It is currently widely anticipated that a theory of quantum gravity will eventually solve the singularity problem. Since the current state of quantum gravity is far from solving this issue, researchers have followed another track: to seek ``regular'' black hole models as a semi-classical alternative to describe nonsingular gravitational objects.

The first ever model of regular black holes was proposed by Bardeen, who introduced a metric without a central singularity by replacing it with a de Sitter-like core \cite{1968qtr87B}. This model was initially phenomenological, not a solution to Einstein's field equations. However, about three decades after Bardeen proposed this model, E. Ayón-Beato and A. García demonstrated that such geometries could be physically obtained by coupling general relativity to nonlinear electrodynamics \cite{Ayon-Beato98,Ayon-Beato00}. In this framework, the magnetic or electric charge of the black hole acts as a source for a nonlinear field that prevents collapse into an infinitesimal point, ensuring the finiteness of curvature invariants such as the Kretschmann scalar at the center of spacetime. Following this breakthrough finding, the field of regular black hole solutions in general relativity coupled to nonlinear electrodynamics expanded with the introduction of the Bronnikov model \cite{Bronnikov:PRL:2000,Bronnikov:PRD:2001}, the Dymnikova model \cite{Dymnikova:2004zc,Dymnikova:2010zz}, which addresses the asymptotic behavior of the spacetimes in the weak-field limit; the Hayward model \cite{Hayward:PhysRevLet:}, which offers a different approach to the core's structure; and the Fan-Wang solutions \cite{Fan:PRD:2016,Bronnikov17,Toshmatov:PRD:comment}, which further generalize these ideas. This formalism has been extensively investigated, leading to the construction of various regular black hole solutions whose physical and geometrical properties have been thoroughly analyzed \cite{Hollenstein:2008hp,Junior:2023ixh,Balart:2014jia,Toshmatov17,Balart:2014cga,Toshmatov:2021fgm,Croney2025,Capozziello:2025ycu,Huang:2025uhv,Tlemissov:2025nnk,Verbin:2024ewl,dePaula:2024yzy,Dolan:2024qqr,Chen:2025aom,Bokulic:2025brf}. In addition to coupling general relativity with nonlinear electrodynamics, several alternative approaches for obtaining regular black hole solutions have also been proposed and studied in the literature \cite{Nicolini:2005vd,Lemos:2011dq,Balakin:2015gpq,Chamseddine:2016ktu,Aros:2019quj,Bueno:2024zsx,Suvorov:2025lar,Bueno:2024dgm,Guerrero:2020uhn,Junior:2023qaq}.

The construction just described has two difficulties that are often left implicit. The first concerns the meaning of the matter sector as a theory. When the metric is prescribed and the Lagrangian is read off from the field equations, the resulting $\L$ is usually written in terms of the mass and charge of the solution one started from. A Lagrangian, however, should define a theory, while the mass and the charge should label its solutions. If $\L$ itself contains $M$ and $Q$, two black holes of different mass belong to two different theories, and a black hole that absorbs a charged particle would leave the theory it came from. The second difficulty is specific to the electric case. For an electrically charged regular solution the invariant $F$ vanishes both at the center and at infinity, so $\L(F)$ necessarily consists of two branches joined at a cusp; Bronnikov has shown that this cannot be avoided when the theory has a Maxwell weak-field limit \cite{Bronnikov:PRL:2000,Bronnikov:PRD:2001,Bronnikov17}. More recently, it has been shown that the mass and the charge of such regular black holes cannot be independent \cite{Bokulic:2025brf}. These issues are connected to each other and to the well-known failure of the first law of thermodynamics for regular black holes. In a recent Letter \cite{letter} we showed, using the family of solutions presented here, that the matter sector can be written as a Hamiltonian density with fixed couplings and no solution parameters, and that the standard first law then holds exactly once the theory is kept fixed. In the present paper, we give the full construction and the details that the Letter could only state: the derivation of the solution family and of its theory, the location of the branch point of $\L(F)$ relative to the horizon, the horizon structure and the energy conditions, the horizon potential in closed form, and the extended first law and Smarr formula with the couplings as thermodynamic variables.

Although, general relativity has been exceptionally successful in weak-field tests, a substantial energy gap remains between gravity and the other fundamental interactions. The fact that gravity is significantly weaker than other natural forces represents a fundamental barrier to the unification of physical theories. To bridge this gap, modern frameworks often propose increasing the gravitational energy scale by including additional spatial dimensions in the spacetime manifold \cite{Arkani-Hamed:1998jmv,Antoniadis:1998ig,Emparan:2008eg}. There is an even greater need for higher dimensions in gravity theories such as the braneworld model, which suggests that our four-dimensional spacetime (the brane) might be a surface within a higher-dimensional spacetime (the bulk) \cite{Maartens:2010ar}. Furthermore, string theory maintains a mathematical requirement for extra dimensions to achieve a consistent theory of quantum gravity \cite{Becker:2006dvp}. We have learned that, over the past three decades, higher-dimensional black holes are much less constrained than four-dimensional black holes. The simplest higher-dimensional black hole spacetime was proposed by Tangherlini \cite{Tangherlini:1963bw}. Studying black holes in higher dimensions is not only a mathematical task, but it is also important for understanding basic properties of gravity, such as uniqueness theorems, the stability of event horizons, and the holographic principle. In this context, various higher-dimensional black hole solutions have been constructed \cite{Myers:1986un,Myers:1986rx,Reall:2002bh,Elvang:2004rt,Galloway:2005mf,Emparan:2006mm,Tomizawa:2006vp,Rogatko:2006gg,Kleihaus:2007kc,Fang:2018rte,Ahmedov:2021ohg,Rahimov:2024hol}. In this paper, we aim to construct the higher-dimensional regular black hole in general relativity coupled to nonlinear electrodynamics. The paper is organized as follows. In Sec.~\ref{sec-background}, we review the background geometry and the corresponding equations of motion. In Sec.~\ref{sec-elec-charged}, we present the formalism for constructing electrically charged black holes in general relativity coupled to nonlinear electrodynamics. In Sec.~\ref{sec-new-solution}, we introduce a new higher-dimensional regular black hole solution and discuss its main properties; there we also recast the matter sector, in the dual Hamiltonian formulation of Sec.~\ref{sec-Pframe}, as a theory with fixed couplings and no solution parameters, determine the solution space that this theory generates, and analyze the branch structure of $\L(F)$. Furthermore, Secs.~\ref{sec-EC} and \ref{sec-thermo} are devoted to the analysis of the energy conditions and thermodynamic properties of the solution, respectively. Finally, in Sec.~\ref{sec-conclusion}, we summarize the main results of this work. Throughout the paper, we adopt the spacelike signature $(-,+,+,\ldots,+)$ and use geometrized units with $G_{D}=1=c$.

\section{Background}\label{sec-background}

The higher-dimensional, spherically symmetric, static spacetime, which we will refer to as the $D$-dimensional metric, is given as
\begin{eqnarray}\label{spacetime}
    ds^2=-f(r)dt^2+\frac{dr^2}{f(r)}+r^2d\Omega_{D-2}^2\ ,
\end{eqnarray}
where $d\Omega_{D-2}^2$ is the metric of the unit sphere $S^{D-2}$, which is given as 
\begin{eqnarray}
    d\Omega_{D-2}^2&=& d\theta_1^{2}+\sin^{2}\theta_1\Big( d\theta_2^{2}+\sin^{2}\theta_2\big(\cdots\nonumber\\
    &&\qquad\qquad\cdots+\sin^{2} \theta_{D-3}\, d\phi^{2} \big) \Big)\\
    &=&\sum_{i=1}^{D-3}\left(\prod_{j=1}^{i-1} \sin^2 \theta_j\right)d\theta_i^2+\left(\prod_{j=1}^{D-3}\sin^2 \theta_j\right)d\phi^2.\nonumber
\end{eqnarray}
Within pure general relativity in a vacuum where $R_{\mu\nu}=0$, the spacetime metric represents Schwarzschild-Tangherlini spacetime with the metric function
\begin{eqnarray}
    f(r)=1-\frac{M}{r^{D-3}}\ ,
\end{eqnarray}
where $M$ is a mass parameter related to the ADM mass of the black hole $M_{\rm ADM}$ as
\begin{eqnarray}\label{MADM-M-rel}
    M_{\rm ADM}=\alpha M\ ,
\end{eqnarray}
where
\begin{eqnarray}\label{alpha}
    \alpha=\frac{(D-2)\Omega_{D-2}}{16\pi}, \quad \Omega_{D-2}=\frac{2\pi^{(D-1)/2}}{\Gamma\left(\frac{D-1}{2}\right)}\ ,
\end{eqnarray}
Here, $\Gamma[x]$ denotes the Gamma function, which for positive integers $x$ satisfies $\Gamma[x]=(x-1)!$, and more generally obeys the relation $\Gamma[x+1]=x\,\Gamma[x]$ with the specific value $\Gamma[1/2]=\sqrt{\pi}$.
\begin{figure}[h]
\centering
\includegraphics[width=0.48\textwidth]{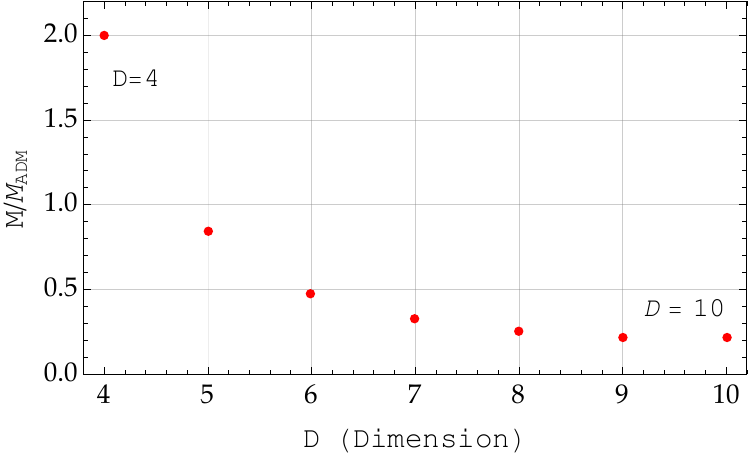}
\caption{The mass parameter as a function of the ADM mass for the spacetime dimension in the range of $D\in[4,10]$.}\label{fig-MADM-M}
\end{figure}
In the case $D=4$, we have $\Omega_{D-2}=4\pi$ and $M=2M_{\rm ADM}$, corresponding to the Schwarzschild spacetime. The expression (\ref{MADM-M-rel}) does not explicitly demonstrate the dependence of the mass parameter on the spacetime dimension. Therefore, in Fig. \ref{fig-MADM-M}, we present this relationship. It shows that the ratio of the mass parameter to the ADM mass decreases with increasing spacetime dimension. 

This spacetime possesses a curvature singularity at its center, as revealed by the Kretschmann scalar, which is given by
\begin{eqnarray}
    K=R_{\mu\nu\alpha\beta}R^{\mu\nu\alpha\beta}=\frac{(D-1)(D-2)^2(D-3)M^2}{r^{2(D-1)}}\ ,
\end{eqnarray}
which diverges as the radial coordinate tends to zero. No coordinate transformation can remove this divergence. Therefore, the Schwarzschild-Tangherlini black hole is singular. To avoid this, we consider a generalized Schwarzschild-Tangherlini-like metric with a radial-coordinate-dependent mass function defined as
\begin{eqnarray}
    f(r)=1-\frac{m(r)}{r^{D-3}}\ .
\end{eqnarray}
where $m(r)$ represents the localized, coordinate-dependent mass function enclosing the total energy within a sphere of radius $r$. This approach is deeply established in the classic Misner-Sharp mass formulation \cite{Misner:1964je}, which has been widely generalized across different spacetimes to investigate the structural, thermodynamic, and evolutionary dynamics of spherically symmetric horizons \cite{Maeda:2007uu,Nielsen:2008kd,Cai:2009qf,Malafarina:2022oka,Panassiti:2025diw,Bonanno:2023rzk}. 

For this case, the Kretschmann scalar takes the following form:
\begin{eqnarray}
    K &=& \frac{1}{r^{2D-2}} \Big\{ \big[ r^2 m'' - 2(D-3)r m'\nonumber\\
    && \qquad\qquad\ \, + (D-2)(D-3)m \big]^2\nonumber\\
    && + 2(D-2)\left( r m' - (D-3)m \right)^2\nonumber\\
    && + 2(D-2)(D-3)m^2 \Big\}\ .
\end{eqnarray}

One can clearly see from the above expression that the Kretschmann scalar tends to a finite value if
\begin{eqnarray}\label{limits}
    &&\lim_{r\rightarrow0}\frac{m(r)}{r^{D-1}}=\text{finite},\nonumber\\
    &&\lim_{r\rightarrow0}\frac{m'(r)}{r^{D-2}}=\text{finite},\\
    &&\lim_{r\rightarrow0}\frac{m''(r)}{r^{D-3}}=\text{finite}.\nonumber
\end{eqnarray}
Thus, by choosing an appropriate form of the mass function that satisfies the conditions (\ref{limits}), one can obtain a regular spacetime. The question of whether this spacetime can also be a solution to Einstein’s field equations is addressed below by coupling general relativity to nonlinear electrodynamics.

The action of general relativity, coupled to nonlinear electrodynamics, is given by
\begin{eqnarray}\label{action}
    S=\frac{1}{16\pi}\int d^Dx\sqrt{-g}\left(R-\L(F)\right)\ ,
\end{eqnarray}
where $g$ is the determinant of the metric tensor $g_{\mu\nu}$, $R$ is the scalar curvature, and $\L$ is the Lagrangian density for the nonlinear electrodynamics field, expressed as a function $\L(F)$ of the electrodynamic field strength $F$, which is the electromagnetic field invariant given by $F_{\mu\nu}F^{\mu\nu}$, where $F_{\mu\nu}$ represents the electromagnetic field tensor. This tensor can be described via a gauge potential as $F_{\mu\nu}=\partial_\mu A_\nu-\partial_\nu A_\mu$. It is noteworthy that, due to its definition, $F_{\mu\nu}$ is an anti-symmetric tensor with only $D(D-1)/2$ independent components.

The action (\ref{action}) results in the following covariant equations of motion, which are the Einstein equations
\begin{eqnarray}\label{Einstein-eq}
    G_{\mu\nu}=8\pi T_{\mu\nu}\ ,
\end{eqnarray}
where the energy-momentum tensor of the nonlinear electrodynamics is given by
\begin{eqnarray}\label{ener-mom-tensor}
    T_{\mu\nu}=\frac{1}{4\pi}\left(\L_FF_{\mu}^{\ \alpha}F_{\nu\alpha}-\frac{1}{4}g_{\mu\nu}\L\right)\ ,
\end{eqnarray}
where $\L_F=\partial\L/\partial F$.

At the same time, nonlinear electrodynamics is governed by Maxwell's equations
\begin{eqnarray}\label{Maxwell-eq}
    \nabla_\mu\left(\L_FF^{\mu\nu}\right)=0\ .
\end{eqnarray}
As we have stated earlier, we adopt the spacetime geometry (\ref{spacetime}) to solve the above given equations of motion. The last component that must be set up to construct the black hole solution is the gauge potential one-form for the electrically charged black hole, given as
\begin{eqnarray}\label{em-potential}
    A=\varphi(r)dt\ ,
\end{eqnarray}
where $\varphi(r)$ is the electric potential, so that the radial electric field is $E(r)=|F_{tr}|=|\varphi'(r)|$ and the invariant (\ref{F-elec}) below is negative, as it must be for a purely electric configuration. 

Thus, as we have set up all necessary equations and conditions, we can proceed to solve the Einstein equations. It is well-known from the symmetry of the Einstein tensor that there are $D(D+1)/2$ numbers of independent components of $G_{\mu\nu}$. Moreover, the symmetry of spacetime also decreases the number of independent components of the Einstein tensor to the two that are given by
\begin{eqnarray}\label{Einstein-tensor}
    &&G_{t}^{\ t}=G_{r}^{\ r}=\frac{D-2}{2r^2}\left[rf'-(D-3)(1-f)\right]\ ,\\
    &&G_{\theta_i}^{\ \theta_i}=G_{\phi}^{\ \phi}\nonumber\\
    &&\qquad=\frac{1}{2} f''+\frac{D-3}{r} f'-\frac{(D-3)(D-4)}{2r^2}\left(1-f\right).\nonumber
\end{eqnarray}
Now, only the right-hand side of the Einstein equation, i.e., the independent components of the energy-momentum tensor, are missing. However, since the structure of the electromagnetic field tensor and, consequently, the energy–momentum tensor differ for the electrically and magnetically charged black hole cases, we present these cases separately in the next sections.

\section{Electrically charged black hole}\label{sec-elec-charged}

In this section, we construct an electrically charged higher-dimensional black hole spacetime in general relativity coupled to nonlinear electrodynamics. For this case, the electromagnetic field 2-form can be written as ${\bf F}_2=\varphi'(r){\bf d}r \wedge{\bf d}t$. As a result, the electromagnetic field invariant takes the following form:
\begin{eqnarray}\label{F-elec}
    F=-2\varphi'(r)^2\ .
\end{eqnarray}
The Maxwell equations (\ref{Maxwell-eq}) produce Gauss's law for the electric field to be 
\begin{eqnarray}\label{elec-pot}
    E(r)=\frac{Q_{\rm ADM}}{r^{D-2}\L_F}\ ,
\end{eqnarray}
where $Q_{\rm ADM}$ is the ADM electric charge. Thus, the electromagnetic field invariant takes the following form:
\begin{eqnarray}\label{F-elec2}
    F=-\frac{2Q_{\rm ADM}^2}{r^{2(D-2)}\L_F^2}\ .
\end{eqnarray}
By substituting the given quantities into the energy-momentum tensor (\ref{ener-mom-tensor}), we obtain the following two independent components:
\begin{eqnarray}\label{EM-tensor-el-char}
    &&T_{t}^{\ t}=T_{r}^{\ r}=-\frac{\L-2F\L_F}{16\pi}\ ,\\
    &&T_{\theta_i}^{\ \theta_i}=T_{\phi}^{\ \phi}=-\frac{\L}{16\pi}.\nonumber
\end{eqnarray}
By solving the Einstein equations (\ref{Einstein-eq}) with the Einstein tensor (\ref{Einstein-tensor}) and the energy-momentum tensor (\ref{EM-tensor-el-char}), we obtain the following expressions for the Lagrangian density of nonlinear electrodynamics and its derivative with respect to the electromagnetic field invariant:
\begin{align}
    &\L=\frac{(D-4) (D-3) (1-f)}{r^2}-\frac{2 (D-3) f'}{r}-f'',\label{L-el-char}\\
    &\L_F=\frac{1}{2\varphi '^2}\left[\frac{(D-4) f'}{2 r}+\frac{(D-3) (1-f)}{r^2}+\frac{f''}{2}\right].\label{LF-el-char}
\end{align}
By combining Maxwell's equations (\ref{elec-pot}), the expression (\ref{LF-el-char}) can be rewritten as
\begin{eqnarray}\label{LF-el-char2}
    \L_F=\frac{4 Q_{\rm ADM}^2 r^{6-2 D}}{r^2 f''+(D-4) r f'-2 (D-3) f+2 D-6}\ .
\end{eqnarray}
In Maxwell's electrodynamics, whose Lagrangian density is a linear function of the electromagnetic field invariant, as $\L=F$, $\L_F=1$, the solutions of both differential equations (\ref{L-el-char}) and (\ref{LF-el-char2}) give the following identical solutions:
\begin{eqnarray}
    f(r)=1-\frac{M}{r^{D-3}}+\frac{Q^2}{r^{2(D-3)}}\ ,
\end{eqnarray}
where $Q$ is the electric charge parameter, of dimension $L^{D-3}$, which is related to the conserved charge $Q_{\rm ADM}$ appearing in Gauss's law (\ref{elec-pot}) by
\begin{eqnarray}\label{QADM-maxwell}
    Q_{\rm ADM}=\sqrt{\frac{(D-2)(D-3)}{2}}\,Q\ .
\end{eqnarray}
We note for later use that $[Q_{\rm ADM}]=L^{D-3}$ in any dimension, as follows directly from (\ref{elec-pot}) together with $[E]=L^{-1}$. This will matter in the next section, where the charge parameter of the regular solutions carries a different dimension and the relation (\ref{QADM-maxwell}) can therefore not be carried over.
Thus, by choosing the Lagrangian density of the electromagnetic field and solving the differential equations with respect to the metric function, one can obtain the corresponding spacetime metric in general relativity coupled to electrodynamics, as we did in the previous example of Maxwell's electrodynamics. However, in this case, we cannot predict that the initially adopted Lagrangian density will produce a regular black hole. On the other hand, any metric function $f(r)$ that satisfies Maxwell's (\ref{elec-pot}) and Einstein's (\ref{L-el-char})-(\ref{LF-el-char2}) equations can be considered a solution of general relativity coupled to electrodynamics. Therefore, we choose the latter method where we select the metric function so that it satisfies the regularity condition (\ref{limits}). Below, we present a new class of regular black hole solutions obtained in this way, and we then show in Secs.~\ref{sec-Pframe}--\ref{sec-solspace} how the matter sector so obtained is promoted to a theory with fixed couplings.

\section{New class of regular black holes}\label{sec-new-solution}

In this section, we propose a new class of electrically charged higher-dimensional regular black holes within the formalism presented in the previous section. Let us choose the following mass function:
\begin{eqnarray}
    m(r) = \frac{M r^{D-1}}{(r^\gamma + Q^\gamma)^{(D-1)/\gamma}}
\end{eqnarray}
which reduces to the following metric function:
\begin{eqnarray}\label{metric-function-generic}
    f(r) = 1 - \frac{M r^2}{(r^\gamma + Q^\gamma)^{(D-1)/\gamma}}\ ,
\end{eqnarray}
where $\gamma$ is a constant positive number through which we generalize several well-known regular black hole spacetimes into one. For example, in the four-dimensional ($D=4$) case, $\gamma=2$ and $\gamma=3$ correspond to the Bardeen and Hayward solutions, respectively, while $\gamma=1$ represents the Maxwellian black hole solution \cite{Fan:PRD:2016,Bronnikov17,Toshmatov:PRD:comment}. The cases corresponding to the other natural values of $\gamma$ represent new types of regular black holes. Thus, the current solution can be considered a generalization of these solutions to higher dimensions as well. 

Before writing down the corresponding matter sector, one remark on dimensions is in order. In (\ref{metric-function-generic}) the charge parameter is added to $r^\gamma$ and therefore has the dimension of a length, $[Q]=L$, in contrast with the Maxwell solution of the previous section, where $[Q]=L^{D-3}$. The relation (\ref{QADM-maxwell}) consequently does not apply here. The conserved charge is instead defined, as always, by the first integral of the Maxwell equations, $Q_{\rm ADM}=r^{D-2}\L_F E$, and the field equations fix it only up to the rescaling $F\to c^{-2}F$, $\L_F\to c^{2}\L_F$, which leaves both the metric and $\L(r)$ untouched and merely redefines the unit of field strength. We fix this freedom once and for all by
\begin{eqnarray}\label{QADM-new}
    Q_{\rm ADM}^2=\frac{(D-1)(\gamma+D-1)}{4}\,M\,Q^{D-3}\ ,
\end{eqnarray}
which has the correct dimension $[Q_{\rm ADM}]=L^{D-3}$ and which, as we verify below, is the normalization for which $\L_F\to1$ in the weak-field regime whenever a Maxwellian limit exists. With this choice, the spacetime solution (\ref{metric-function-generic}) corresponds to the following nonlinear electrodynamics Lagrangian density
\begin{align}
    &\L=(D-1) M Q^{\gamma } \frac{(D-2) Q^{\gamma }-(\gamma +1) r^{\gamma }}{\left(r^{\gamma }+Q^{\gamma }\right)^{(2 \gamma +D-1)/\gamma}}\ ,\label{L-density}\\
    &\L_F=\frac{\left(r^{\gamma }+Q^{\gamma }\right)^{(2 \gamma +D-1)/\gamma}}{Q^{\gamma-D+3}\,r^{2D+\gamma-4}}\ .
\end{align}
By using the relationship $F'=\L'/\L_F$ where the prime denotes the derivative with respect to the radial coordinate, we find that the electromagnetic field invariant is 
\begin{eqnarray}\label{F-invariant}
    F=-\frac{(D-1)(\gamma +D-1) M\, Q^{2 \gamma -D+3}\, r^{2 (\gamma +D-2)}}{2\left(r^{\gamma }+Q^{\gamma }\right)^{2(2 \gamma +D-1)/\gamma}}.
\end{eqnarray}
In the weak-field regime where $r\rightarrow\infty$ (or $r^\gamma\gg Q^\gamma$), the spacetime metric is asymptotically flat and behaves like
\begin{eqnarray}
    f(r)=1-\frac{M}{r^{D-3}}+\frac{M(D-1)Q^\gamma}{\gamma r^{D+\gamma-3}}+{\cal O}\!\left(r^{3-D-2\gamma}\right).
\end{eqnarray}
The first two terms of the above expression represent the Schwarzschild-Tangherlini spacetime, while the third term represents the contribution of the electromagnetic field to the spacetime at spatial infinity. The positive sign of the third term indicates that the charge parameter $Q$ acts as a ``screening'' mechanism, effectively weakening the gravitational attraction compared to the Schwarzschild-Tangherlini black hole of the same mass $M$. The electromagnetic invariant (\ref{F-invariant}) behaves as 
\begin{eqnarray}\label{F-weak}
    F \sim r^{-2(\gamma+1)}.
\end{eqnarray}
On the other hand, the Maxwell electrodynamics in $D$-dimensions behaves as
\begin{eqnarray}
    F \sim r^{-2(D-2)}.
\end{eqnarray}
The standard Maxwell fall-off is recovered only for $\gamma=D-3$, while deviations from this value correspond to non-Maxwellian nonlinear electrodynamics.

What about the Lagrangian density in the weak-field regime? To answer this question, we find the asymptotic behavior of the Lagrangian density (\ref{L-density}). In this limit, it behaves like
\begin{eqnarray}
    \L\sim r^{-(\gamma+D-1)}\ ,
\end{eqnarray}
which, in terms of (\ref{F-weak}), can be written as
\begin{eqnarray}
    \L_{\rm asymp}\sim |F|^{\frac{\gamma+D-1}{2(\gamma+1)}}\ .
\end{eqnarray}
Consequently, only in the case where $\gamma=D-3$ does the Lagrangian density of nonlinear electrodynamics exhibit Maxwellian behavior in the weak-field regime. This further supports the findings in \cite{Fan:PRD:2016} that the Hayward ($\gamma=3$) and Bardeen ($\gamma=2$) models of regular black holes fail to recover a Maxwellian limit in four-dimensional spacetime, whereas the $\gamma=1$ model maintains the correct limit. This particular drawback in certain nonlinear electrodynamic models was previously highlighted in \cite{Bronnikov:PRL:2000,Burinskii:2002pz,Toshmatov:2019gxg}. Interestingly, in higher-dimensional spacetimes, $\gamma=1$ no longer yields a Maxwellian limit; rather, the limit is achieved by models with higher values of $\gamma$. For instance, the Bardeen regular black hole in five dimensions and the Hayward regular black hole in six dimensions both possess the correct Maxwellian limits.

Let us now consider the behavior of spacetime and nonlinear electrodynamics in the strong-field regime, $r\rightarrow0$ (or $r^\gamma\ll Q^\gamma$). As it approaches the regular center, the spacetime with the metric function (\ref{metric-function-generic}) exhibits an asymptotic de Sitter behavior given by
\begin{eqnarray}
f(r)=1-\frac{M}{Q^{D-1}}r^2+{\cal O}(r^{\gamma+2})\ .
\end{eqnarray}
Concurrently, the electromagnetic field invariant (\ref{F-invariant}) behaves as
\begin{eqnarray}
F\sim r^{2(\gamma+D-2)}\ .
\end{eqnarray}
Given that $\gamma$ is a positive constant and $D\geq4$, the exponent is strictly positive, meaning that the electromagnetic field invariant vanishes at the center of the spacetime ($\lim_{r\to0}F = 0$). In this deep core limit ($r\rightarrow0$), the Lagrangian density (\ref{L-density}) can be expressed as a function of the field invariant via the asymptotic expansion:
\begin{eqnarray}
\L_{\rm core}\sim\L_0+c_\ast |F|^{\frac{\gamma}{2(\gamma+D-2)}}\ ,
\end{eqnarray}
where the coefficient $c_\ast$ is a constant determined entirely by the parameters of the spacetime. Due to its highly cumbersome form, its explicit expression is omitted here. The leading-order term, $\L_0$, acts as a constant vacuum contribution, that is, as an effective cosmological constant $\Lambda_{\rm eff}=\L_0/2$ in the normalization of the action (\ref{action}), which balances out the singularity. It is given by
\begin{eqnarray}
\L_0=\frac{(D-1)(D-2)M}{Q^{D-1}}\ .
\end{eqnarray}
Thus, $\L_0$ perfectly matches and generates the regular de Sitter behavior of the spacetime in the central core region.

It is worth noting that obtaining a single-valued, explicit analytical expression for the Lagrangian $\L(F)$ valid over the entire spacetime manifold is obstructed by the non-monotonicity of the field invariant $F(r)$. Since $F$ vanishes both at the center and at infinity, it must reach an extremum at some radius $r_{\rm crit}$ where $dF/dr=0$. At this turning point $F(r)$ cannot be inverted, and the Lagrangian splits into an inner branch ($\L_{\rm core}$) for $r<r_{\rm crit}$ and an outer branch ($\L_{\rm asymp}$) for $r>r_{\rm crit}$. This happens for every $\gamma>0$, and it is a general feature of regular electric solutions \cite{Bronnikov:PRL:2000,Bronnikov17}; we return to it in Sec.~\ref{sec-branch}. The problem is absent in the dual description, which we now introduce and which we use to define the theory.

\subsection{Dual Hamiltonian formulation in the $P$ framework}\label{sec-Pframe}

To avoid the complications introduced by the turning point of $F(r)$, we switch to the dual $P$ framework \cite{Ayon-Beato98,Bronnikov:PRL:2000,Salazar:1987ap}. In this dual representation the field content is carried by the auxiliary antisymmetric tensor
\begin{equation}
P_{\mu\nu} = \L_F F_{\mu\nu},
\end{equation}
and by its invariant $P \equiv P_{\mu\nu} P^{\mu\nu}$, which we normalize exactly as $F$, so that $P=F$ in the Maxwell limit $\L_F=1$. For the static, spherically symmetric configuration under consideration, the equations of motion (\ref{Maxwell-eq}) read $\nabla_\mu P^{\mu\nu} = 0$ and are integrated by Gauss's law (\ref{elec-pot}), which fixes the single nonvanishing component to be $P_{tr} = -Q_{\rm ADM}/r^{D-2}$. Consequently,
\begin{equation}\label{P-of-r}
P(r) = -\frac{2 Q_{\rm ADM}^2}{r^{2(D-2)}}\ ,
\end{equation}
which is negative, as $F$ is for a purely electric field. In contrast to $F(r)$, the dual invariant is strictly monotonic over the entire domain $r \in (0, \infty)$, which permits an exact, single-valued global inversion free of branch ambiguities,
\begin{equation}\label{r-of-P}
r(P) = \left( \frac{2 Q_{\rm ADM}^2}{|P|} \right)^{\frac{1}{2(D-2)}}.
\end{equation}
The field theory in the dual picture is specified by the Hamiltonian density obtained through the Legendre transformation \cite{Salazar:1987ap}
\begin{eqnarray}\label{Legendre}
    \mathcal{H} = 2 F \L_F - \L\ ,
\end{eqnarray}
whose derivative satisfies $\mathcal{H}_P \equiv d\mathcal{H}/dP = 1/\L_F$, and which is inverted by
\begin{equation}\label{inv-Legendre}
F_{\mu\nu} = \mathcal{H}_P \, P_{\mu\nu}, \qquad \L(F) = 2 P \mathcal{H}_P - \mathcal{H}\ .
\end{equation}
Comparison of (\ref{Legendre}) with the energy-momentum tensor (\ref{EM-tensor-el-char}) shows that the Hamiltonian density is proportional to the energy density of the source, with a negative proportionality constant,
\begin{equation}\label{H-equals-rho}
\mathcal{H}(r) = 16\pi\, T^{t}_{\ t}(r) = -16\pi \rho(r)\ ,
\end{equation}
a relation that is easily checked in the Maxwell case, where $\L=F$ gives $\mathcal{H}=F=-2E^2$ and $\rho=E^2/8\pi$. Substituting the inversion (\ref{r-of-P}) into the energy density of the present solution, we obtain the explicit, globally single-valued closed form
\begin{equation}\label{H-of-P-raw}
\mathcal{H}(P) = -\frac{(D-2)(D-1) M Q^\gamma}{\left[\left( \dfrac{2 Q_{\rm ADM}^2}{|P|} \right)^{\frac{\gamma}{2(D-2)}}+Q^\gamma \right]^{\frac{D-1+\gamma}{\gamma}}}\ .
\end{equation}
Since both $\mathcal{H}=-16\pi\rho$ and $P$ increase with $r$, one has $\mathcal{H}_P>0$ everywhere, in agreement with $\L_F>0$. Finally, the physical radial electric field $E(r) = |F_{tr}|$ and the electrostatic potential are recovered from (\ref{inv-Legendre}) as
\begin{equation}\label{E-and-phi}
E(r) = \mathcal{H}_P \frac{Q_{\rm ADM}}{r^{D-2}}, \qquad \varphi(r) = \int_{r}^{\infty} E(r')\, dr'\ ,
\end{equation}
the potential being normalized so that it vanishes at spatial infinity.

\subsection{The matter sector as a theory with fixed couplings}\label{sec-theory}

Expression (\ref{H-of-P-raw}) is single-valued, but it still contains $M$ and $Q$, which belong to one particular solution. A theory should be defined independently of its solutions, with the mass and the charge entering only as integration constants. We now show that this is possible here, and that the appearance of $M$ and $Q$ in (\ref{H-of-P-raw}) is only a consequence of how the solution was generated.

Using the normalization (\ref{QADM-new}), we introduce the two dimensionful constants
\begin{eqnarray}\label{couplings}
    &&\sigma\equiv\frac{(D-2)(D-1)M}{Q^{D-1}}\ ,\nonumber\\
    &&\beta\equiv\frac{2Q_{\rm ADM}^2}{Q^{2(D-2)}}=\frac{(D-1)(\gamma+D-1)M}{2Q^{D-1}}\ ,
\end{eqnarray}
both of the same dimension as $P$ and $F$. Using (\ref{P-of-r}), one has $r^\gamma/Q^\gamma=(\beta/|P|)^{\gamma/[2(D-2)]}$, and (\ref{H-of-P-raw}) becomes
\begin{equation}\label{H-of-P}
\mathcal{H}(P) = -\sigma\left[1+\left(\dfrac{\beta}{|P|}\right)^{\frac{\gamma}{2(D-2)}}\right]^{-\frac{\gamma+D-1}{\gamma}}\ .
\end{equation}
The function (\ref{H-of-P}) contains no solution parameters. It is fixed by the number $\gamma$ and the two coupling constants $\sigma$ and $\beta$, just as Born-Infeld electrodynamics is fixed by its critical field. The theory is the action (\ref{action}) with the matter sector given by the Legendre dual of (\ref{H-of-P}).

The two couplings have a simple meaning. In the strong-field regime $|P|\to\infty$ one finds $\mathcal{H}\to-\sigma$ and $\L\to\sigma$, so $\sigma=\L_0$ is the vacuum energy of the core that produces the de Sitter behavior at small $r$. In the weak-field regime $|P|\to0$ one finds
\begin{equation}\label{H-weak}
\mathcal{H}(P)\simeq -\sigma\left(\frac{|P|}{\beta}\right)^{\frac{\gamma+D-1}{2(D-2)}}\ .
\end{equation}
The Maxwellian behavior $\mathcal{H}\propto P$ is recovered if and only if the exponent equals unity, that is if and only if
\begin{equation}
\gamma=D-3\ ,
\end{equation}
in agreement with the condition found above from $\L(F)$. In that case $\mathcal{H}\to(\sigma/\beta)P$, so the ratio of the two couplings sets the strength of the Maxwell coupling at infinity. For the present family this ratio is fixed by (\ref{couplings}) to be
\begin{eqnarray}\label{sigma-over-beta}
    \frac{\sigma}{\beta}=\frac{2(D-2)}{\gamma+D-1}\ ,
\end{eqnarray}
which equals unity exactly when $\gamma=D-3$. With the normalization (\ref{QADM-new}) the Maxwell limit is therefore $\mathcal{H}\to P$, or $\L_F\to1$, with no further rescaling. Note that the condition $\gamma=D-3$ was found earlier from the outer branch of $\L(F)$; here it follows from the single-valued function $\mathcal{H}(P)$, so it is a property of the theory and not of a branch.

\subsection{Solution space of the fixed theory}\label{sec-solspace}

With the theory fixed, we now ask what its static, spherically symmetric, electrically charged solutions are. Gauss's law gives (\ref{P-of-r}) with $Q_{\rm ADM}$ now a free integration constant, which we trade for the length
\begin{equation}\label{lambda-def}
    \lambda\equiv\left(\frac{2Q_{\rm ADM}^2}{\beta}\right)^{\frac{1}{2(D-2)}}\ ,
\end{equation}
in terms of which (\ref{H-of-P}) gives the radial energy density profile
\begin{equation}\label{rho-fixed-theory}
    \rho(r)=-\frac{\mathcal{H}}{16\pi}=\frac{\sigma}{16\pi\left[1+(r/\lambda)^{\gamma}\right]^{\frac{\gamma+D-1}{\gamma}}}\ .
\end{equation}
Writing the metric function as $f=1-\hat m(r)/r^{D-3}$, the $tt$ Einstein equation reduces to $\hat m'=16\pi\rho\,r^{D-2}/(D-2)$, whose general solution is
\begin{equation}\label{mhat-general}
    \hat m(r)=m_0+\frac{\sigma\lambda^{D-1}}{D-2}\,\mathcal{J}\!\left(\frac{r}{\lambda}\right),
\end{equation}
with $m_0$ the second integration constant and
\begin{equation}\label{Jfun}
    \mathcal{J}(x)=\int_0^{x}\frac{y^{D-2}\,dy}{\left(1+y^{\gamma}\right)^{\frac{\gamma+D-1}{\gamma}}}=\frac{x^{D-1}}{(D-1)\left(1+x^{\gamma}\right)^{\frac{D-1}{\gamma}}}\ ,
\end{equation}
so that $\mathcal{J}(\infty)=1/(D-1)$. The fixed theory therefore has a two-parameter family of solutions, labeled by $(m_0,Q_{\rm ADM})$. For $m_0=0$, Eq.~(\ref{Jfun}) returns exactly the metric function (\ref{metric-function-generic}), with the identifications given in (\ref{mass-charge}) below.

Expanding (\ref{mhat-general}) near the origin gives
\begin{equation}
    f(r)=1-\frac{m_0}{r^{D-3}}-\frac{\sigma\, r^2}{(D-1)(D-2)}+{\cal O}(r^{\gamma+2})\ ,
\end{equation}
so regularity of the center is a condition on the solution, $m_0=0$, and not on the theory. Imposing it, the regular solutions form a one-parameter family labeled by the charge alone,
\begin{equation}\label{mass-charge}
    Q=\lambda\ ,\qquad M=\frac{\sigma\,\lambda^{D-1}}{(D-2)(D-1)}\ ,
\end{equation}
that is, $M\propto Q_{\rm ADM}^{(D-1)/(D-2)}$. Inserting (\ref{couplings}) one checks that (\ref{mass-charge}) gives back the parameters of the solution (\ref{metric-function-generic}). In four dimensions the relation becomes $M\propto Q_{\rm ADM}^{3/2}$, which is precisely the mass-charge relation recently proved to be unavoidable for regular black holes sourced by nonlinear electrodynamics \cite{Bokulic:2025brf}; this agreement is a useful check of the present formulation.

Three consequences follow, and they answer the difficulties raised in the Introduction. First, the mass and the charge are true integration constants of a fixed theory. Second, a black hole of this theory that absorbs a small charged particle remains a solution of the same theory: the couplings do not change, only the integration constants do. Third, the final state will in general have $m_0\neq0$ and will therefore be singular at the center. Regularity is a special property of one subfamily of solutions and is not preserved by accretion. This is a physical limitation of regular black holes sourced by nonlinear electrodynamics, not an inconsistency of the construction, and it agrees with the general results of Ref.~\cite{Bokulic:2025brf}.

\subsection{Event horizon}

\begin{figure*}[ht]
\includegraphics[width=0.33\textwidth]{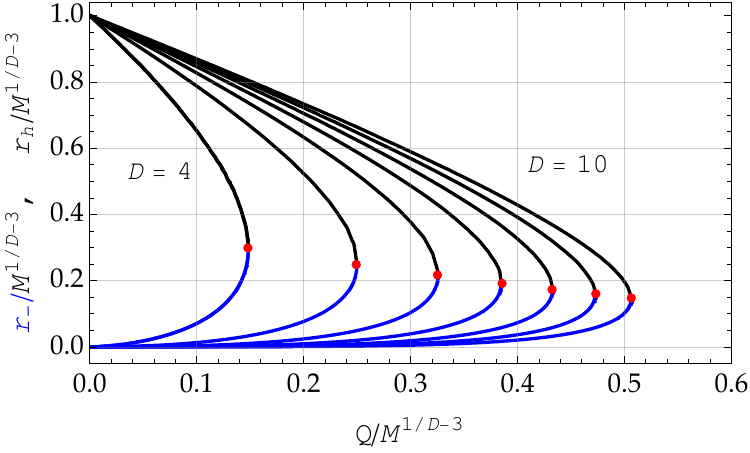}
\includegraphics[width=0.33\textwidth]{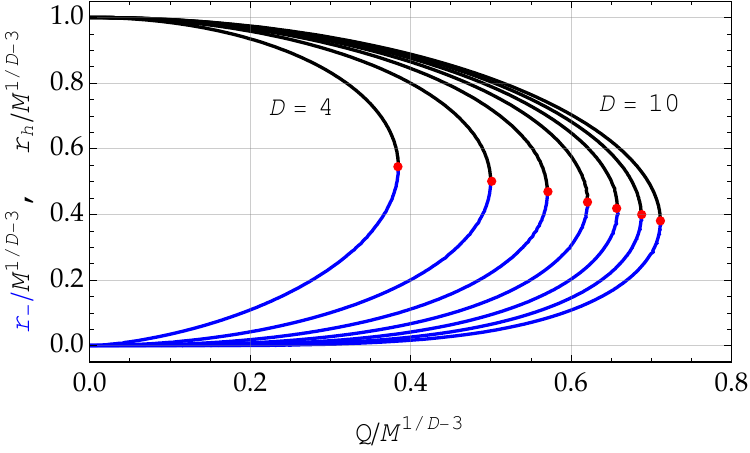}
\includegraphics[width=0.33\textwidth]{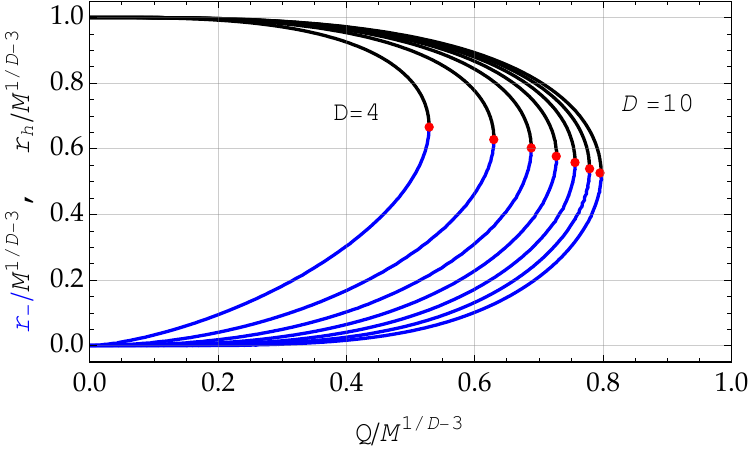}
\caption{\label{fig-horizon} Dependence of the radius of the coordinate singularity (the event horizon -- black, the inner horizon -- blue) on the charge parameter for the dimension of the spacetime $D\in[4,10]$, (from left to right) $\gamma=1,\ 2,\ 3$ cases. Here the red dots indicate the possible maximum value of the charge parameter and minimum value of the event horizon radius of the black hole, i.e., the extremely charged black hole, determined by (\ref{rmin-qmax}).}
\end{figure*}

The event horizon of the black hole is determined by solving the equation $f(r)=0$ with respect to the radial coordinate. This equation admits at most two positive solutions, which represent the inner horizon, $r_-$, and the outer (event) horizon, $r_{\rm h}$. Unfortunately, this equation cannot be solved analytically. Therefore, we solve it numerically and present the dependence of the event horizon radius on the charge parameter, $\gamma$, and the dimension of the spacetime in Fig. \ref{fig-horizon}.
Thus, one can see in Fig. \ref{fig-horizon} that the regular black holes under consideration have two horizons until the charge parameter reaches the maximum value, at which point both the inner and outer horizons merge, resulting in the black holes having the smallest possible horizon radii (red dots in Fig. \ref{fig-horizon}). The radius of the extreme event horizon and the corresponding value of the charge parameter are determined by
\begin{align}\label{rmin-qmax}
    &r_{\rm h,min}=\left(\frac{2}{D-1}\right)^{(D-1)/(\gamma(D-3))}M^{1/(D-3)}\ ,\\
    &Q_{\rm max}=(D-3)^{1/\gamma}\left[\frac{4}{(D-1)^{D-1}}\right]^{1/(\gamma(D-3))}M^{1/(D-3)}.\nonumber
\end{align}
It is often more convenient to express the same extremal configuration with the charge parameter, rather than the mass, held fixed. Requiring the two roots of $f(r)=0$ to coincide is equivalent to the vanishing of the surface gravity, and it gives the minimal horizon radius and the corresponding minimal mass parameter as
\begin{align}\label{r-min}
    &r_{\ast}=\left(\frac{2}{D-3}\right)^{1/\gamma}Q\ ,\\
    &M_\ast=\left(\frac{2}{D-3}\right)^{-2/\gamma} \left(\frac{D-1}{D-3}\right)^{(D-1)/\gamma}Q^{D-3}\ ,\nonumber
\end{align}
which are readily checked to be equivalent to (\ref{rmin-qmax}). Beyond this value of the charge parameter, the spacetime represents a no-horizon object. At a fixed mass, the maximum admissible value of the charge parameter grows with the spacetime dimension while the minimum event horizon radius decreases; the latter trend persists at fixed charge, as (\ref{r-min}) shows directly. Even if the event horizon radius of the black hole for arbitrary values of the spacetime parameters cannot be solved, in a weakly charged black hole approximation, we can find it as
\begin{eqnarray}\label{r-hor-approx}
    r_{\rm h}\approx M^{1/(D-3)}\left[1-\frac{(D-1)Q^{\gamma}}{(D-3)\gamma}M^{-\gamma/(D-3)}\right]\ .
\end{eqnarray}
The first term of the right-hand side of (\ref{r-hor-approx}) represents the event horizon radius of the Schwarzschild-Tangherlini black hole. It also shows that for small values of the charge parameter, in the case of $\gamma=1$, the event horizon radius decreases as a linear function of the charge parameter, while in the cases of $\gamma=$ 2 and 3, it decreases as quadratic and cubic functions of the charge parameter, respectively.

\subsection{Branch structure of $\L(F)$ and its causal location}\label{sec-branch}

We now return to the two branches of $\L(F)$ and ask where the branch point lies. Differentiating (\ref{F-invariant}), $dF/dr$ vanishes at the single positive root
\begin{equation}\label{rcrit}
    r_{\rm crit}=\left(\frac{\gamma+D-2}{\gamma+1}\right)^{1/\gamma}Q\ .
\end{equation}
At this radius $\L_F$ is finite and positive, but $\L_{FF}=\L_F'(r)/F'(r)$ diverges, so that the two branches of $\L(F)$ meet at a cusp. This is the behavior identified by Bronnikov as the obstruction to a globally defined electric Lagrangian \cite{Bronnikov:PRL:2000,Bronnikov17}. In the dual description nothing special happens: since $P'(r)\neq0$, the function $\mathcal{H}(P)$ is smooth at $r_{\rm crit}$. The cusp is thus a feature of the choice of $F$ as the variable, and the present model is best described as a nonlinear electrodynamics defined in the Hamiltonian form, whose Lagrangian form has two branches.

This also explains why the family with $\gamma=D-3$ does not contradict the no-go theorems of Refs.~\cite{Bronnikov:PRL:2000,Bokulic:2025brf}, which forbid regular electric black holes in theories with a Maxwell weak-field limit. Those theorems assume a single-valued $\L(F)$; here the single-valued object is $\mathcal{H}(P)$, and $\L(F)$ has two branches. The price is the cusp itself, so it matters whether the cusp is visible from outside the black hole.

\begin{figure}[h]
\centering
\includegraphics[width=0.48\textwidth]{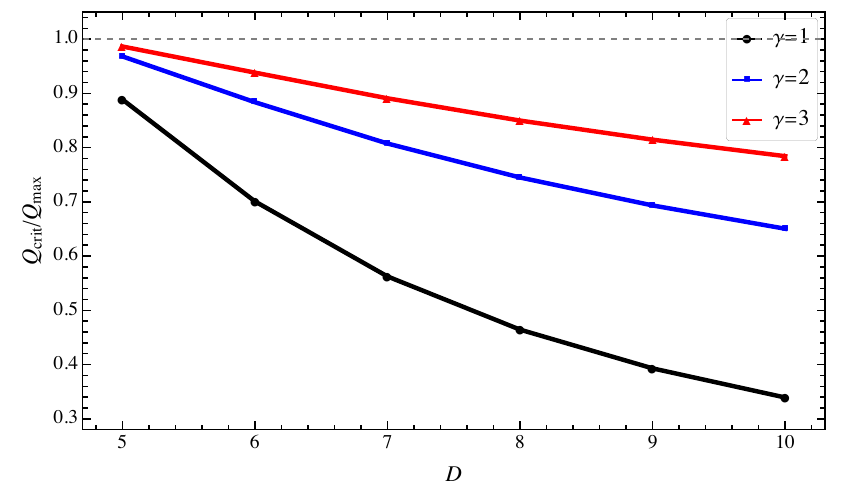}
\caption{The ratio $Q_{\rm crit}/Q_{\rm max}$ as a function of the spacetime dimension for $\gamma=1$ (black), $\gamma=2$ (blue) and $\gamma=3$ (red). For charges below $Q_{\rm crit}$ the branch point (\ref{rcrit}) of $\L(F)$ is hidden behind the event horizon; between each curve and the dashed line it lies outside the horizon. In $D=4$ the branch point is always hidden, so the plot starts at $D=5$.}\label{fig-branch}
\end{figure}

Comparing (\ref{rcrit}) with the extremal radius (\ref{r-min}), one finds that $r_{\rm crit}\geq r_\ast$ if and only if $(\gamma+D-2)(D-3)\geq2(\gamma+1)$. This fails for every $\gamma>0$ when $D=4$ and holds for every $\gamma>0$ when $D\geq5$. In four dimensions, therefore, the cusp is always inside the event horizon, whatever the mass and the charge. For $D\geq5$ the cusp is inside the horizon only if the black hole is massive enough for a given charge, or equivalently only if the charge is below a critical value $Q_{\rm crit}$, defined by $r_{\rm h}=r_{\rm crit}$. One finds $Q_{\rm crit}<Q_{\rm max}$ for all $D\geq5$, so there is always a range of near-extremal charges, $Q_{\rm crit}<Q\leq Q_{\rm max}$, for which the branch point lies outside the horizon. Fig.~\ref{fig-branch} shows $Q_{\rm crit}/Q_{\rm max}$ as a function of the dimension. The range widens quickly with $D$, because $r_{\rm crit}$ grows with $D$ while $r_\ast$ shrinks, and it narrows as $\gamma$ increases.
The conclusion is twofold. If the Hamiltonian formulation (\ref{H-of-P}) is taken as the definition of the theory, as we do here and as is standard for electric nonlinear electrodynamics \cite{Salazar:1987ap,Ayon-Beato98}, the model is well defined for all values of the parameters. If instead one insists on a Lagrangian $\L(F)$, then for $D\geq5$ the charge should be kept below $Q_{\rm crit}$, so that the cusp stays hidden from external observers. This difference between four and higher dimensions does not seem to have been noticed before.

\section{Energy conditions}\label{sec-EC}

To ensure the solution represents a physically realistic source, in this section, we investigate the energy conditions by introducing the energy density $\rho$, radial pressure $p_r$, and tangential pressure $p_t$ as follows:
\begin{align}\label{rho-p}
    &\rho(r)=-T_{t}^{\ t}\ , \quad p_r(r)=T_{r}^{\ r}\ ,\nonumber\\
    &p_t(r)=T_{\theta_i}^{\ \theta_i}=T_{\phi}^{\ \phi}\ .
\end{align}
These quantities take the following forms for the spacetime metric (\ref{spacetime}) with the metric function (\ref{metric-function-generic}): 
\begin{align}\label{rho-pr-pt}
    &\rho(r)=-p_r(r)=\frac{(D-2) (D-1) M Q^{\gamma } }{16\pi\left(r^{\gamma }+Q^{\gamma }\right)^{(\gamma +D-1)/\gamma }}\ ,\\
    &p_t(r)=\frac{(D-1) M Q^{\gamma } \left[(\gamma +1) r^{\gamma }-(D-2) Q^{\gamma }\right]}{16\pi\left(r^{\gamma }+Q^{\gamma }\right)^{(2\gamma +D-1)/\gamma }},
\end{align}
The energy density, radial pressure, and tangential pressure are all finite at the center of the spacetime, where they approach the isotropic de Sitter values
\begin{eqnarray}
    \rho_0=-p_{r0}=-p_{t0}=\frac{(D-2) (D-1) M}{16\pi Q^{D-1}}\ .
\end{eqnarray}
In the figure below, Fig. \ref{fig-distr}, we present the qualitative behavior of the radial profiles of the normalized energy density, radial pressure, and tangential pressure.
\begin{figure*}[ht]
\includegraphics[width=0.33\textwidth]{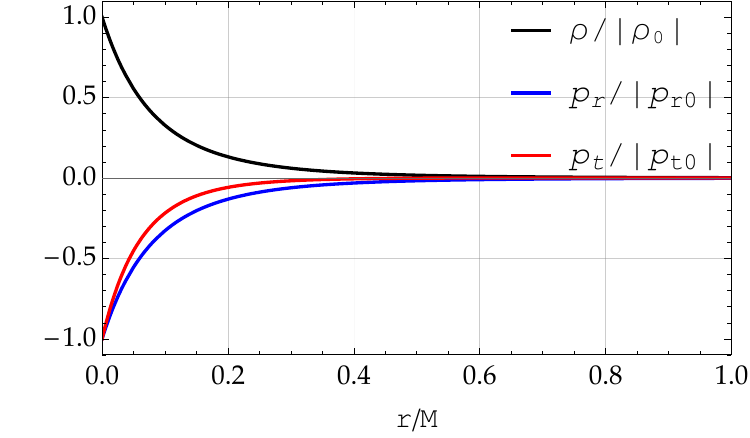}
\includegraphics[width=0.33\textwidth]{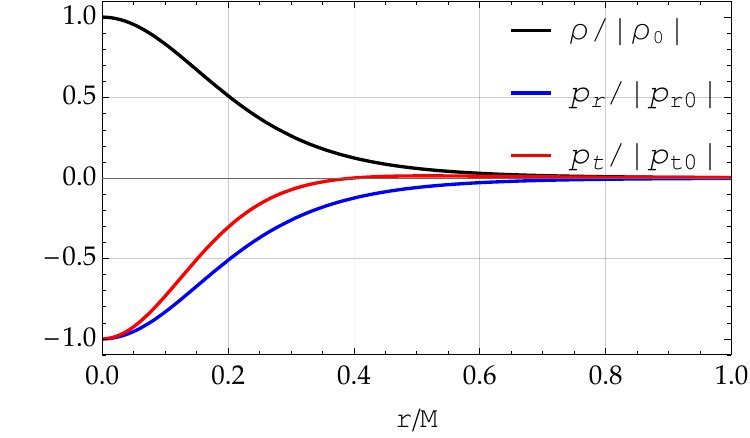}
\includegraphics[width=0.33\textwidth]{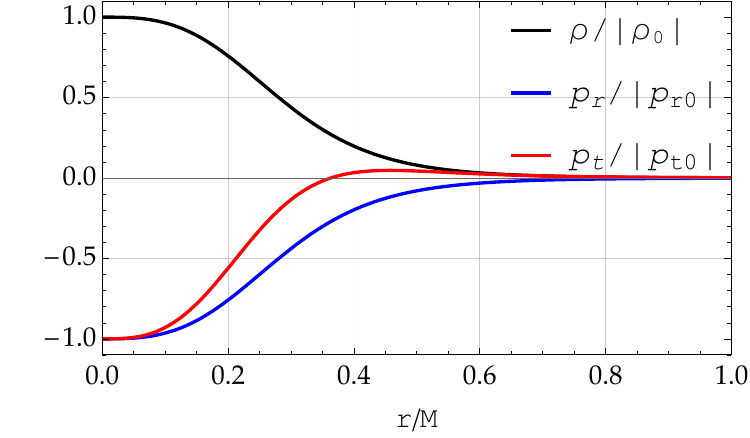}
\caption{\label{fig-distr} The radial profiles of the energy density, radial pressure and tangential pressure for (from left to right) $\gamma=1,\ 2,\ 3$ cases. Here we set $D=5$, and $Q=0.4M$.}
\end{figure*}
One can see from Fig. \ref{fig-distr} that all of these quantities asymptotically tend to zero at spatial infinity. The energy density is positive everywhere, $\rho>0$, and the radial pressure, being its exact opposite, is everywhere negative. The tangential pressure, by contrast, behaves differently from the other two components: it is negative close to the center of spacetime, where the source is de Sitter-like, and becomes positive at larger values of the radial coordinate. Thus, the tangential pressure vanishes at the radius
\begin{eqnarray}\label{pt=0}
    r_{t}=\left(\frac{D-2}{\gamma+1}\right)^{1/\gamma}Q\ .
\end{eqnarray}
From (\ref{pt=0}) and Fig. \ref{fig-distr}, it is clear that for greater values of the parameter $\gamma$ and lower-dimensional spacetimes, the negative tangential pressure region is narrower. Below, we investigate the energy conditions separately for the spacetime under consideration.

\subsection{Null energy condition}

The null energy condition (NEC) states that for any future-directed null vector $k^\mu$, the following relation must hold:
\begin{eqnarray}\label{nec}
    T_{\mu\nu}k^\mu k^\nu\geq0\ ,
\end{eqnarray}
The physical interpretation of the NEC is that the energy density measured by an observer following a null geodesic is always non-negative. If we rewrite (\ref{nec}) in terms of the spacetime metric (\ref{spacetime}) and the energy-momentum tensor (\ref{rho-p}), we obtain the following:
\begin{eqnarray}\label{nec1}
    \rho+p_j\geq0, \quad j=r,\theta_i,\phi\ .
\end{eqnarray}
From (\ref{rho-pr-pt}), it is obvious that $\rho+p_r=0$, and 
\begin{eqnarray}
    \rho+p_t=\frac{(D-1)MQ^\gamma r^\gamma(D+\gamma-1)}{16\pi(r^\gamma+Q^\gamma)^{(D+2\gamma-1)/\gamma}}\ ,
\end{eqnarray}
which is zero at the center of spacetime and positive elsewhere. Thus, the NEC is always satisfied in this spacetime for both black hole and no-horizon cases.

\subsection{Weak energy condition}

In the weak energy condition (WEC), the energy-momentum tensor must satisfy the following inequality for every future-pointing timelike vector $u^\mu$:
\begin{eqnarray}\label{wec}
    T_{\mu\nu}u^\mu u^\nu\geq0\ ,
\end{eqnarray}
The physical interpretation of the WEC is that the energy density measured by an observer moving along the time-like geodesic with four-velocity $u^\mu$ is never negative. In terms of the spacetime metric (\ref{spacetime}) and the energy-momentum tensor (\ref{rho-p}), the WEC (\ref{wec}) takes the following form:
\begin{eqnarray}\label{wec1}
    \rho\geq0\quad {\rm and}\quad  \rho+p_j\geq0, \quad j=r,\theta_i,\phi\ .
\end{eqnarray}
Comparing the null and weak energy conditions, we observe that the second inequality of the WEC in Eq. (\ref{wec1}) coincides with the NEC condition given in Eq. (\ref{nec1}). This is consistent with the well-known hierarchy of classical energy conditions, according to which the WEC implies the NEC; i.e., if the WEC holds, the NEC is automatically satisfied,
\begin{eqnarray}\label{wec-nec}
    {\rm WEC}\Rightarrow{\rm NEC}\ .
\end{eqnarray}
As we have already mentioned, the energy density $\rho$ is always positive, and the NEC is always satisfied. Therefore, the WEC is also satisfied in this spacetime.

\subsection{Strong energy condition}

While the NEC and WEC focus on the non-negativity of energy density, the strong energy condition (SEC) focuses on the trace of the energy-momentum tensor. The mathematical formulation of the SEC is given by
\begin{eqnarray}\label{sec}
    \left(T_{\mu\nu}-\frac{1}{D-2}g_{\mu\nu}T\right)u^\mu u^\nu\geq0\ ,
\end{eqnarray}
where $T=T_\mu^\mu$ is the trace of the energy-momentum tensor. If we rewrite this condition in terms of the Einstein equations (\ref{Einstein-eq}), it takes the following form:
\begin{eqnarray}
    R_{\mu\nu}u^\mu u^\nu\geq0\ .
\end{eqnarray}
Thus, in other words, the SEC is a principle in general relativity that states that the gravitational behavior of all physical matter should be attractive. In terms of the current spacetime metric and the energy-momentum tensor, it can be rewritten as
\begin{eqnarray}\label{sec1}
    {\rm SEC}
    =(D-3)\rho+\sum_{j=r}^{\phi}p_j\geq0\ .
\end{eqnarray}
Since
\begin{align}
&{\rm SEC}
=(D-3)\rho+p_r+(D-2)p_t=\\
&\frac{(D-1)(D-2)MQ^\gamma}
{16\pi(r^\gamma+Q^\gamma)^{(2\gamma+D-1)/\gamma}}
\left[(\gamma+D-3)r^\gamma-2Q^\gamma
\right]\geq0,\nonumber
\end{align}
the SEC is violated in the region
\begin{equation}\label{sec-viol}
r<r_{\rm SEC}\equiv
\left(\frac{2}{\gamma+D-3}\right)^{1/\gamma}Q.
\end{equation}
Therefore, the SEC is violated only in the vicinity of the regular center, while it is recovered outside this region. Such a violation is typical for regular black holes and is associated with their de Sitter-like behavior in the core of the spacetime \cite{Zaslavskii:2010qz,Rodrigues:2018bdc}. The de Sitter-like behavior indicates the gravitational repulsive nature of the matter, which is obviously against the SEC.

\subsection{Dominant energy condition}

Finally, we consider the most restrictive and physically intuitive energy condition, the DEC, which in turn implies the WEC,
\begin{eqnarray}\label{dec-wec}
    {\rm DEC}\Rightarrow{\rm WEC}\ .
\end{eqnarray}
While the WEC ensures that the energy density is non-negative (\ref{wec}), the DEC extends it further by adding the following condition to the WEC:
\begin{eqnarray}
    J^\mu=-T^{\mu}_{\ \nu} v^\nu\ ,
\end{eqnarray}
where $J^\mu$ is the energy-momentum flux measured by an observer with four-velocity $v^\mu$. Here, the DEC implies the $J^\mu$ to be timelike or lightlike (null) as
\begin{eqnarray}
    g_{\mu\nu}J^\mu J^\nu\leq0\ .
\end{eqnarray}
This condition indicates that the speed of the energy-momentum flow does not exceed the speed of light. Thus, the DEC requires not only the WEC to hold, but it also ensures that energy-momentum cannot flow faster than the speed of light. The DEC can be rewritten as follows in terms of the components of the energy-momentum tensor:
\begin{eqnarray}\label{dec1}
    \rho\geq0,\quad \rho+p_j\geq0, \quad \rho-p_j\geq0,\quad j=r,\theta_i,\phi\ .
\end{eqnarray}
In (\ref{dec1}), the first two conditions represent the WEC, while the third one guarantees relativistic causality. As we have considered in the previous subsections, the WEC is satisfied for the current spacetime. Let us write the last condition of (\ref{dec1}) in terms of the energy density and tangential pressure (\ref{rho-pr-pt}) as
\begin{align}\label{dec2}
    &\rho-p_r=2\rho>0\ ,\\
    &\rho-p_t=\frac{(D-1)MQ^\gamma \left[r^\gamma(D-\gamma-3)+2(D-2)Q^\gamma\right]}{16\pi(r^\gamma+ Q^\gamma)^{(D+2\gamma-1)/\gamma}}.\nonumber
\end{align}
From the above expressions, it is obvious that the DEC is always satisfied everywhere in spacetime if $\gamma\leq D-3$. However, in the opposite case, $\gamma>D-3$, the DEC is violated in the region $r\in(r_{0},\infty)$, where
\begin{eqnarray}\label{r0-dec}
    r_0=\left[\frac{2(D-2)}{\gamma-D+3}\right]^{1/\gamma}Q\ .
\end{eqnarray}
\begin{figure}[h]
\centering
\includegraphics[width=0.48\textwidth]{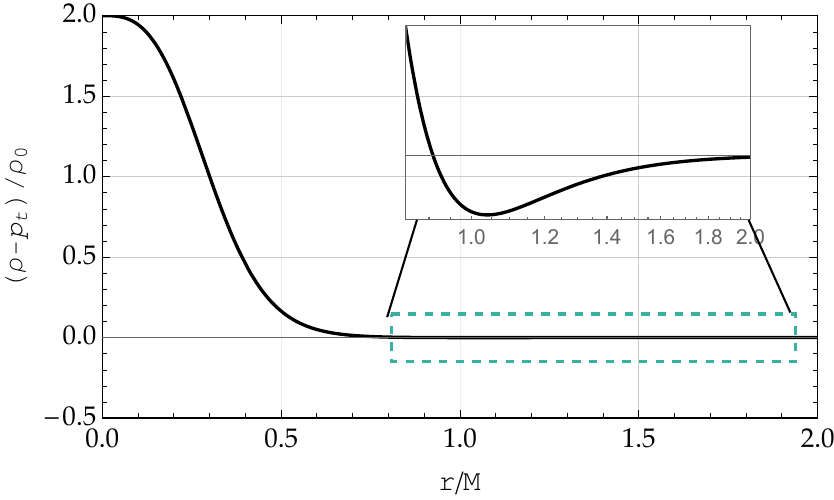}
\caption{The radial profile of $\rho-p_t$ for the DEC violated case. Here we set $\gamma=3$, $D=5$, $Q=0.5$, $M=1$.}\label{fig-dec-violation}
\end{figure}
In Fig. \ref{fig-dec-violation}, we present the DEC violation. In summary, regarding the energy conditions, the present solution of general relativity minimally coupled to nonlinear electrodynamics satisfies both the NEC and the WEC throughout the entire spacetime manifold. The SEC is fulfilled only in the region $r\geq r_{\rm SEC}$, where $r_{\rm SEC}$ is given by (\ref{sec-viol}), while the dominant energy condition (DEC) is satisfied everywhere in spacetime if and only if $\gamma \leq D - 3$. For $\gamma > D - 3$, the DEC holds exclusively within the interval $r \in [0, r_0]$, where $r_0$ is given by (\ref{r0-dec}).

\section{Thermodynamics}\label{sec-thermo}

Since the event horizon radius of the black hole cannot be found analytically in the generic form of the metric function (\ref{metric-function-generic}), we describe the ADM mass of the black hole via the relationship (\ref{MADM-M-rel}) in terms of the event horizon as
\begin{eqnarray}
    M_{\rm ADM}=\alpha\frac{(r_{\rm h}^\gamma+Q^\gamma)^{(D-1)/\gamma}}{r_{\rm h}^2}
\end{eqnarray}
and the surface gravity
\begin{eqnarray}
    \kappa=\frac{M_{\rm ADM} r_{\rm h}\left[(D-3)r_{\rm h}^{\gamma}-2 Q^{\gamma}\right]}{2\alpha\left(r_{\rm h}^{\gamma}+Q^{\gamma}\right)^{(\gamma+D-1)/\gamma}}
\end{eqnarray}
whose vanishing reproduces the minimal (extremal) horizon radius $r_\ast$ and the minimal mass parameter $M_\ast$ already given in (\ref{r-min}). The expression for the minimal horizon radius of the black hole (\ref{r-min}) also confirms the results presented in Fig. \ref{fig-horizon}, which show that with increasing spacetime dimension, this possible minimal size of the black hole decreases. The Hawking temperature of the black hole is determined via the surface gravity of the black hole horizon as $T_{\rm h}=\kappa/(2\pi)$, which turns out
\begin{eqnarray}\label{Hawking-temp}
    T_{\rm h}=\frac{1}{4\pi r_{\rm h}}\left(\frac{(D-1)r_{\rm h}^{\gamma}}{r_{\rm h}^{\gamma}+Q^{\gamma}}-2\right)\ .
\end{eqnarray}
In the limit where the charge vanishes ($Q \to 0$), this expression smoothly recovers the standard Hawking temperature of the $D$-dimensional Schwarzschild-Tangherlini black hole, $T_{\rm h}=(D-3)/(4\pi r_{\rm h})$. Below in Fig. \ref{fig-Hawking}, we present the dependence of the Hawking temperature of the higher-dimensional regular black hole on the event horizon radius and other spacetime parameters of the black hole. 
\begin{figure*}[ht]
\includegraphics[width=0.45\textwidth]{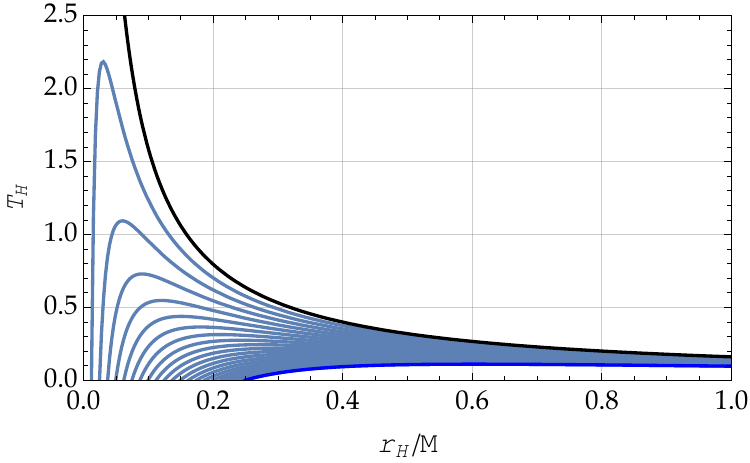}
\includegraphics[width=0.45\textwidth]{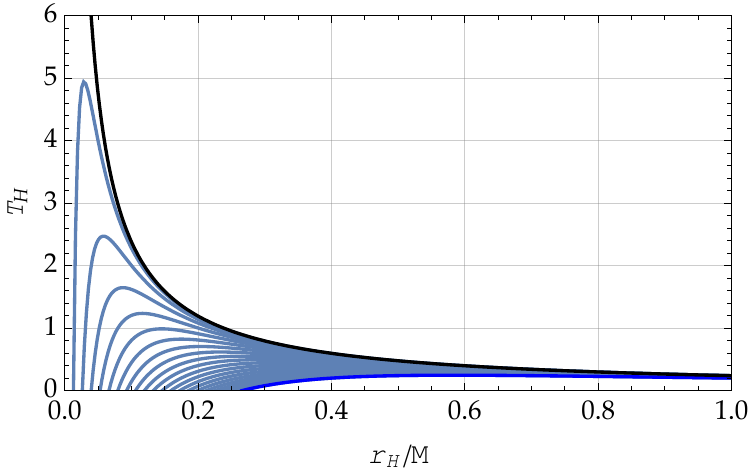}
\caption{\label{fig-Hawking} Dependence of the Hawking temperature of a higher-dimensional regular black hole on the event horizon radius and other spacetime parameters. The charge varies from $Q=0$ (black curve) to $Q=Q_{\rm max}$ (blue curve) in steps of 0.05. The left and right panels correspond to $D=5$, $\gamma=1$, and $D=6$, $\gamma=2$, respectively.}
\end{figure*}
One can see in Fig. \ref{fig-Hawking} that, at a fixed horizon radius, the Hawking temperature grows with the spacetime dimension but is reduced by the charge parameter, falling to zero for the minimal (extremal) black hole. The entropy of the black hole is defined by the Bekenstein-Hawking area law, $S=A_{D-2}/4$, which reduces to
\begin{eqnarray}\label{entropy}
    S=\frac{\Omega_{D-2}r_{\rm h}^{D-2}}{4}\ ,
\end{eqnarray}
and it satisfies the relation
\begin{eqnarray}
    \frac{\partial S}{\partial r_{\rm h}}=\frac{(D-2)S}{r_{\rm h}}=4\pi\alpha r_{\rm h}^{D-3}\ .
\end{eqnarray}
The entropy in the weakly charged black hole regime takes the following form:
\begin{eqnarray}
    S \approx S_0 \left[ 1 - \frac{(D-2)(D-1)}{\gamma(D-3)} \left( \frac{Q}{M^{1/(D-3)}} \right)^{\gamma}\right],
\end{eqnarray}
where $S_0$ is the entropy of the Schwarzschild-Tangherlini black hole ($Q=0$). It indicates that the entropy of the higher-dimensional charged regular black hole decreases due to the charge parameter relative to the Schwarzschild-Tangherlini black hole.

Let us now consider the variation of the ADM mass along the family (\ref{metric-function-generic}), taking the charge parameter as a thermodynamic variable together with its conjugate $\Phi_{\rm h}^{\rm (br)}$,
\begin{eqnarray}\label{thermo-1}
    dM_{\rm ADM}=T_{\rm thermo} dS+\Phi_{\rm h}^{\rm (br)} dQ\ ,
\end{eqnarray}
where
\begin{eqnarray}\label{partials}
   &&T_{\rm thermo}= \left(\frac{\partial M_{\rm ADM}}{\partial S}\right)_{Q},\quad \Phi_{\rm h}^{\rm (br)}= \left(\frac{\partial M_{\rm ADM}}{\partial Q}\right)_{S}.
\end{eqnarray}
Since the mass parameter is not an explicit function of the entropy, the Hawking temperature in (\ref{partials}) can be rewritten as
\begin{eqnarray}
    T_{\rm thermo}= \left(\frac{\partial M_{\rm ADM}}{\partial S}\right)_{Q}=\left[\frac{\left(\partial M_{\rm ADM}/\partial r_{\rm h}\right)}{\left(\partial S/{\partial r_{\rm h}}\right)}\right]_{Q}\ .
\end{eqnarray}
where
\begin{align}
    &\left(\frac{\partial M_{\rm ADM}}{\partial r_{\rm h}}\right)_{Q}=\frac{\alpha M}{r_{\rm h}}\left[\frac{(D-1)r_{\rm h}^\gamma}{r_{\rm h}^\gamma+Q^\gamma}-2\right]\ ,\\
    &\frac{\partial r_{\rm h}}{\partial S}=\frac{1}{4\pi\alpha r_{\rm h}^{D-3}}\ .
\end{align}
Thus, the temperature defined by this variation becomes
\begin{eqnarray}\label{T-thermo}
    T_{\rm thermo}=\frac{M_{\rm ADM}}{4\pi\alpha r_{\rm h}^{D-2}}\left[\frac{(D-1)r_{\rm h}^\gamma}{r_{\rm h}^\gamma + Q^\gamma}-2\right],
\end{eqnarray}
while the corresponding conjugate potential reads
\begin{eqnarray}\label{Phi-pot}
    \Phi_{\rm h}^{\rm (br)}\equiv\left(\frac{\partial M_{\rm ADM}}{\partial Q}\right)_{S}=\frac{(D-1) Q^{\gamma-1}}{r_{\rm h}^\gamma + Q^\gamma}M_{\rm ADM}\ ,
\end{eqnarray}
where the superscript ``br'' stands for ``branch'' and anticipates the fact, established below, that this derivative is not the electrostatic potential of the horizon.

A comparison of Eqs.~(\ref{Hawking-temp}) and (\ref{T-thermo}) immediately reveals a discrepancy that is well known in this context: the temperature obtained by differentiating the mass differs from the geometric Hawking temperature by the dimensionless factor
\begin{eqnarray}\label{Xi-def}
    \Xi\equiv\frac{T_{\rm thermo}}{T_{\rm h}}=\frac{M}{r_{\rm h}^{D-3}}=\frac{M_{\rm ADM}}{m_{\rm h}}\ ,\quad m_{\rm h}\equiv\alpha\, r_{\rm h}^{D-3},
\end{eqnarray}
where $m_{\rm h}$ is the mass inside the horizon. Since $\Xi>1$ whenever the charge is nonzero, (\ref{thermo-1}) is not the standard first law with the Hawking temperature, and if one keeps both the Hawking temperature and the area law the first law seems to fail. The same happens for the regular black holes of Refs. \cite{Ayon-Beato98,Ayon-Beato00,Bronnikov:PRL:2000,Bronnikov:PRD:2001,Dymnikova:2004zc,Dymnikova:2010zz}, and it has led to proposals in which the entropy is changed or the internal energy is redefined \cite{Ma:2014qma,rasheed2003,Breton:2004qa,Zhang:2016ilt}.

Before accepting such a change, one should ask what is actually varied in (\ref{thermo-1}). The energy density (\ref{rho-pr-pt}) contains $M$ and $Q$, so changing $M$ at fixed $Q$ changes the matter itself. In the language of Sec.~\ref{sec-theory}, the couplings (\ref{couplings}) depend on $M$ and $Q$, so the derivative $(\partial M_{\rm ADM}/\partial S)_Q$ compares black holes of two different theories. A first law, however, relates neighboring solutions of one theory. With the theory held fixed, there is no discrepancy at all, as shown in Ref. \cite{letter}, we recall the result and then develop it further.

\subsection{The first law at fixed theory}\label{sec:first-law-fixed}

In Ref. \cite{letter}, we showed that when the couplings $(\sigma,\beta,\gamma)$ are kept fixed and only the integration constants $(m_0,Q_{\rm ADM})$ of (\ref{mhat-general}) are varied, the first law holds in its standard form,
\begin{equation}\label{first-law-fixed}
    \delta M_{\rm ADM}=T_{\rm h}\,\delta S+\Phi_{\rm h}\,\delta Q_{\rm e}\ ,
\end{equation}
with the Hawking temperature, the unmodified area entropy, and
\begin{equation}\label{Qe-Phi}
    Q_{\rm e}\equiv\frac{\Omega_{D-2}}{4\pi}Q_{\rm ADM},\quad \Phi_{\rm h}=\varphi(r_{\rm h})=\int_{r_{\rm h}}^{\infty}E(r)\,dr\ ,
\end{equation}
where $E(r)$ and $\varphi(r)$ are given in (\ref{E-and-phi}). The proof is short and uses only that the energy density (\ref{rho-fixed-theory}) does not contain $m_0$: one varies the horizon condition $\hat m(r_{\rm h})=r_{\rm h}^{D-3}$ and compares with $T_{\rm h}\delta S$. We refer to Ref. \cite{letter} for the derivation and only use the result here. In four dimensions $Q_{\rm e}=Q_{\rm ADM}$ and (\ref{first-law-fixed}) has the usual Reissner-Nordstr\"om form.

On the regular branch, where $\lambda=Q$, the potential term takes the closed form
\begin{equation}\label{PhiQ-closed}
    \Phi_{\rm h}Q_{\rm e}=\frac{D-1}{D-2}\left[M_{\rm ADM}-m_{\rm h}+\frac{m_{\rm h}\,Q^{\gamma}}{r_{\rm h}^{\gamma}+Q^{\gamma}}\right]\ ,
\end{equation}
which differs from $\Phi_{\rm h}^{\rm (br)}Q$ of (\ref{Phi-pot}); the latter is a derivative along the regular branch, not the horizon potential. Thus, for the regular black holes constructed here the first law holds in its standard form, and neither a modified internal energy nor an extra work term is needed.

\subsection{Variations through the space of theories}\label{sec:theory-space}

It remains to understand the factor $\Xi$, which is a real feature of the family (\ref{metric-function-generic}) even though it is not a failure of the first law. When the regular solutions are labeled by $M$ and $Q$ independently, they form a curve that moves through the space of theories, because by (\ref{couplings}) both couplings depend on $M$ and $Q$.

The natural framework is then an extended first law in which the couplings are treated as thermodynamic variables, as is done for the cosmological constant in extended black hole thermodynamics \cite{Kastor:2009wy} and for the couplings of nonlinear electrodynamics in the Smarr formula \cite{Gulin:2017ycu,Bokuli2021,Bokulic:2025brf}. Adding the variations of the couplings to (\ref{first-law-fixed}) gives
\begin{equation}\label{extended-first-law}
    \delta M_{\rm ADM}=T_{\rm h}\,\delta S+\Phi_{\rm h}\,\delta Q_{\rm e}+\Psi_\sigma\,\delta\sigma+\Psi_\beta\,\delta\beta\ ,
\end{equation}
where $\Psi_\sigma$ and $\Psi_\beta$ are the partial derivatives of $M_{\rm ADM}$ with respect to $\sigma$ and $\beta$ at fixed $S$ and $Q_{\rm e}$. Both follow in closed form from (\ref{mhat-general}):
\begin{equation}\label{couplings-conjugates}
    \sigma\Psi_\sigma=M_{\rm ADM}-m_{\rm h}\ ,\qquad \beta\Psi_\beta=-\frac{1}{2}\Phi_{\rm h}Q_{\rm e}\ .
\end{equation}
The first relation has a clear meaning: the quantity conjugate to the amplitude coupling is simply the energy of the electromagnetic field outside the horizon, $M_{\rm ADM}-m_{\rm h}=\Omega_{D-2}\int_{r_{\rm h}}^{\infty}\rho\,r^{D-2}dr$. It is a work-like term, but one that is derived rather than postulated.

The Smarr formula now follows from Euler's theorem. Counting length dimensions, $M_{\rm ADM}$, $S$ and $Q_{\rm e}$ scale as $L^{D-3}$, $L^{D-2}$ and $L^{D-3}$, while $\sigma$ and $\beta$ scale as $L^{-2}$, so that
\begin{eqnarray}\label{Smarr-full}
    (D-3)M_{\rm ADM}&=&(D-2)T_{\rm h}S+(D-3)\Phi_{\rm h}Q_{\rm e}\nonumber\\
    &&-\,2\sigma\Psi_\sigma-2\beta\Psi_\beta\ .
\end{eqnarray}
Inserting (\ref{couplings-conjugates}) reduces this to
\begin{equation}\label{Smarr-compact}
    (D-1)M_{\rm ADM}=(D-2)T_{\rm h}S+(D-2)\Phi_{\rm h}Q_{\rm e}+2m_{\rm h}\ ,
\end{equation}
which, using (\ref{Hawking-temp}) and (\ref{PhiQ-closed}), is easily checked to be an identity for the solution (\ref{metric-function-generic}). Eq. (\ref{Smarr-full}) is a physical Smarr formula: it contains the Hawking temperature, the area entropy, the horizon potential, and matter terms that were derived. By contrast, the relation
\begin{equation}\label{Euler-identity}
    (D-3)M_{\rm ADM}=(D-2)T_{\rm thermo}S+\Phi_{\rm h}^{\rm (br)}Q
\end{equation}
obtained from (\ref{thermo-1}) is only the Euler identity of the function $M_{\rm ADM}(S,Q)$ along the regular branch. It is exact and is a useful check of (\ref{T-thermo}) and (\ref{Phi-pot}), but it contains $T_{\rm thermo}$ instead of the Hawking temperature and the branch derivative instead of the horizon potential. Replacing $T_{\rm thermo}$ by $T_{\rm h}$ in (\ref{Euler-identity}) breaks the identity, which is why a Smarr formula written naively with the Hawking temperature does not close.

Finally, we note how the factor $\Xi$ arises. Restricting (\ref{extended-first-law}) to the regular branch, where $\sigma$, $\beta$ and $Q_{\rm e}$ are given functions of $M$ and $Q$ through (\ref{QADM-new}) and (\ref{couplings}), one finds $\Xi^{-1}dM_{\rm ADM}=T_{\rm h}dS+\Phi\,dQ$ with $Q\Phi=(D-2)\Phi_{\rm h}Q_{\rm e}-(D-1)(M_{\rm ADM}-m_{\rm h})$. The prefactor $1/\Xi=m_{\rm h}/M_{\rm ADM}$ is the correction factor of the internal-energy scheme of Ref.~\cite{Ma:2014qma}; it is the fraction of the ADM mass inside the horizon, left over after the coupling terms have been absorbed, and it is not a new internal energy, since the one-form $\Xi^{-1}dM_{\rm ADM}$ is not closed. The details are given in Ref.~\cite{letter}.

\section{Conclusion}\label{sec-conclusion}

It is well known that the vacuum black hole solutions of general relativity possess a curvature singularity at the spacetime center. E. Ay{\'o}n-Beato and A. {Garc{\'{\i}}a} \cite{Ayon-Beato98} demonstrated that this singularity can be eliminated by coupling general relativity to nonlinear electrodynamics. Over the past two decades, numerous regular black hole configurations have been developed within this framework. In this work, we presented a formalism for constructing electrically charged, higher-dimensional regular black holes in general relativity sourced by nonlinear electrodynamics. The proposed formalism allows one either to generate higher-dimensional black hole solutions starting from a chosen nonlinear electrodynamics Lagrangian density or, alternatively, to prescribe suitable metric functions and thereby ensure that the resulting spacetime satisfies the Einstein equations.

Within this framework, we proposed a new family of electrically charged, higher-dimensional regular black hole solutions that not only extends the pioneering Hayward and Bardeen models but also yields an additional, distinct class of regular black hole configurations. Instead of simply reading off a Lagrangian from the prescribed metric, we placed the construction on a firmer footing in the dual $P$ framework. The dual invariant is monotonic, so the Hamiltonian density is single-valued everywhere; more importantly, it can be written as (\ref{H-of-P}), a function of $P$ alone with two coupling constants and the exponent $\gamma$, and with no solution parameters. The mass and the charge of a black hole are thus restored to their proper role as integration constants. The fixed theory has a two-parameter family of solutions, and the center is regular only when one of the two integration constants vanishes. The regular solutions therefore form a one-parameter family in which the mass is tied to the charge; in four dimensions this reduces to the relation recently shown to be unavoidable for such black holes \cite{Bokulic:2025brf}. Two consequences follow. A black hole that absorbs a charged particle stays within the same theory, so the construction is consistent; but the final state is in general singular, so regularity is not preserved by accretion.

We also studied the two branches of the Lagrangian $\L(F)$, which cannot be avoided for regular electric solutions \cite{Bronnikov:PRL:2000,Bronnikov17}. The branch point lies at the radius (\ref{rcrit}), where $\L_{FF}$ diverges while the dual description stays smooth. Comparing this radius with that of the extremal horizon gives a result that seems not to have been noticed before: in four dimensions the branch point is always inside the event horizon, whereas for $D\geq5$ it lies outside the horizon for near-extremal charges. This also explains why the $\gamma=D-3$ family, which is electric, regular and Maxwellian at large distances, does not contradict the no-go theorems of Refs.~\cite{Bronnikov:PRL:2000,Bokulic:2025brf}: those theorems assume a single-valued $\L(F)$, whereas here the single-valued object is $\mathcal{H}(P)$.

Our analysis showed that, within our class of regular black hole models, obtaining the Maxwellian limit of nonlinear electrodynamics in the weak-field regime strictly depends on how the model parameter relates to the spacetime dimension, namely through the condition $\gamma=D-3$. This finding aligns with previous results by reaffirming that although the $\gamma=1$ model correctly reproduces the Maxwellian limit in four dimensions, the Hayward ($\gamma=3$) and Bardeen ($\gamma=2$) models do not. Interestingly, the $\gamma=1$ configuration ceases to provide a Maxwellian limit as the number of dimensions increases; instead, larger values of $\gamma$ are needed to recover linear electrodynamics. As a result, models such as the five-dimensional Bardeen black hole and the six-dimensional Hayward black hole are found to exhibit the appropriate Maxwellian limits in their respective spacetimes.

Like those earlier regular solutions, the models we introduce display the same characteristic horizon structure. The presence of the charge endows the black hole with two horizons, an inner one and an outer (event) one. Increasing the electric charge reduces the radius of the event horizon while increasing that of the inner horizon, which signals a weakening of the gravitational attraction of the black hole. When the charge parameter reaches a certain threshold value, the two horizons coincide at the possible minimal radius of the event horizon. Beyond this value of the charge parameter, the horizons vanish, and the object does not represent a black hole. With increasing spacetime dimension, the range of values for the charge parameter increases for the black hole, while the possible minimum radius of the event horizon decreases.

On the other hand, the extra dimension in spacetime enhances the gravitational attraction of the black hole. Moreover, we have presented that the proposed new class of black holes always satisfies the NEC and WEC, but for particular values of the spacetime parameters, the SEC and DEC can be violated. To be more precise, the SEC is violated close to the center of the black hole, in the region $r<r_{\rm SEC}$, where $r_{\rm SEC}$ is given by (\ref{sec-viol}), and the DEC is violated in the region $r>r_0$, where $r_0$ is given by (\ref{r0-dec}), only if $\gamma>D-3$. By comparing the expressions of $r_{\rm SEC}$ (\ref{sec-viol}) and the minimum event horizon radius $r_\ast$ (\ref{r-min}), we find that $r_{\rm SEC}<r_\ast\leq r_{\rm h}$; i.e., the SEC violation always occurs inside the black hole. 

Furthermore, we have derived the surface gravity, the Hawking temperature and the entropy, and we have revisited the temperature mismatch that is usually reported for regular black holes. Differentiating the ADM mass along the regular family at fixed charge gives a temperature larger than the Hawking temperature by the factor $\Xi=M/r_{\rm h}^{D-3}$, and this has often been read as a breakdown of the first law requiring a modified entropy or a redefined internal energy. In Ref.~\cite{letter} we showed that this reading is not necessary: the factor $\Xi$ appears because such a variation changes the couplings of the theory, and once the theory is kept fixed the first law holds exactly in its standard form (\ref{first-law-fixed}). Here we have worked out the consequences for the present family. The horizon potential has the closed form (\ref{PhiQ-closed}); treating the couplings as thermodynamic variables gives the extended first law (\ref{extended-first-law}), whose coupling conjugates are the field energy outside the horizon and one half of the electric term; and Euler's theorem then gives the Smarr formula (\ref{Smarr-full}), which reduces to the compact identity (\ref{Smarr-compact}). The Euler identity (\ref{Euler-identity}) along the regular branch, written with $T_{\rm thermo}$, remains a useful check but should not be mistaken for a Smarr formula.

We are planning to extend these frameworks to analyze the dynamic stability, quasinormal modes, and shadow characteristics of these spacetimes under gravitational and electromagnetic perturbations.

\section*{acknowledgments}
This research was partially supported by the Ministry of Higher Education, Science, and Innovation of the Republic of Uzbekistan under Grants No. FL-9524114991 and No. FL-10425067111. B.A. and C.Y. were supported by the National Natural Science Foundation of China (NSFC) under Grant No. U2541210. 

In the preparation of this work, the author used Claude (Anthropic) to search the literature, and to draft and edit the text under the author's direction. All derivations, numerical results and statements were obtained by the authors, who take full responsibility for the content of the manuscript.

\bibliography{references}

@book{Hawking:1973uf,
    author = "Hawking, Stephen W. and Ellis, George F. R.",
    title = "{The Large Scale Structure of Space-Time}",
    doi = "10.1017/9781009253161",
    isbn = "978-1-009-25316-1, 978-1-009-25315-4, 978-0-521-20016-5, 978-0-521-09906-6, 978-0-511-82630-6, 978-0-521-09906-6",
    publisher = "Cambridge University Press",
    series = "Cambridge Monographs on Mathematical Physics",
    month = "2",
    year = "2023"
}

@INPROCEEDINGS{1968qtr87B,
       author = {{Bardeen}, James},
        title = "{Non-singular general relativistic gravitational collapse}",
    booktitle = {Proceedings of the 5th International Conference on Gravitation and the Theory of Relativity},
         year = 1968,
        month = sep,
        pages = {87},
       adsurl = {https://ui.adsabs.harvard.edu/abs/1968qtr..conf...87B}
}

@article{PhysRevLett.14.57,
  title = {Gravitational Collapse and Space-Time Singularities},
  author = {Penrose, Roger},
  journal = {Phys. Rev. Lett.},
  volume = {14},
  issue = {3},
  pages = {57--59},
  numpages = {0},
  year = {1965},
  month = {Jan},
  publisher = {American Physical Society},
  doi = {10.1103/PhysRevLett.14.57},
  url = {https://link.aps.org/doi/10.1103/PhysRevLett.14.57}
}

@article{Kastor:2009wy,
    author = "Kastor, David and Ray, Sourya and Traschen, Jennie",
    title = "{Enthalpy and the Mechanics of AdS Black Holes}",
    eprint = "0904.2765",
    archivePrefix = "arXiv",
    primaryClass = "hep-th",
    doi = "10.1088/0264-9381/26/19/195011",
    journal = "Class. Quant. Grav.",
    volume = "26",
    pages = "195011",
    year = "2009"
}

@article{Rodrigues:2018bdc,
    author = "Rodrigues, Manuel E. and de Sousa Silva, Marcos V.",
    title = "{Bardeen Regular Black Hole With an Electric Source}",
    eprint = "1802.05095",
    archivePrefix = "arXiv",
    primaryClass = "gr-qc",
    doi = "10.1088/1475-7516/2018/06/025",
    journal = "JCAP",
    volume = "06",
    pages = "025",
    year = "2018"
}

@article{Zaslavskii:2010qz,
    author = "Zaslavskii, O. B.",
    title = "{Regular black holes and energy conditions}",
    eprint = "1004.2362",
    archivePrefix = "arXiv",
    primaryClass = "gr-qc",
    doi = "10.1016/j.physletb.2010.04.031",
    journal = "Phys. Lett. B",
    volume = "688",
    pages = "278--280",
    year = "2010"
}

@article{Hayward:PhysRevLet:,
    author = "Hayward, Sean A.",
    title = "{Formation and evaporation of regular black holes}",
    eprint = "gr-qc/0506126",
    archivePrefix = "arXiv",
    doi = "10.1103/PhysRevLett.96.031103",
    journal = "Phys. Rev. Lett.",
    volume = "96",
    pages = "031103",
    year = "2006"
}

@book{Becker:2006dvp,
    author = "Becker, K. and Becker, M. and Schwarz, J. H.",
    title = "{String theory and M-theory: A modern introduction}",
    doi = "10.1017/CBO9780511816086",
    isbn = "978-0-511-25486-4, 978-0-521-86069-7, 978-0-511-81608-6",
    publisher = "Cambridge University Press",
    month = "12",
    year = "2006"
}

@article{Maartens:2010ar,
    author = "Maartens, Roy and Koyama, Kazuya",
    title = "{Brane-World Gravity}",
    eprint = "1004.3962",
    archivePrefix = "arXiv",
    primaryClass = "hep-th",
    doi = "10.12942/lrr-2010-5",
    journal = "Living Rev. Rel.",
    volume = "13",
    pages = "5",
    year = "2010"
}

@article{Tangherlini:1963bw,
    author = "Tangherlini, F. R.",
    title = "{Schwarzschild field in n dimensions and the dimensionality of space problem}",
    doi = "10.1007/BF02784569",
    journal = "Nuovo Cim.",
    volume = "27",
    pages = "636--651",
    year = "1963"
}

@article{Emparan:2008eg,
    author = "Emparan, Roberto and Reall, Harvey S.",
    title = "{Black Holes in Higher Dimensions}",
    eprint = "0801.3471",
    archivePrefix = "arXiv",
    primaryClass = "hep-th",
    doi = "10.12942/lrr-2008-6",
    journal = "Living Rev. Rel.",
    volume = "11",
    pages = "6",
    year = "2008"
}

@article{Antoniadis:1998ig,
    author = "Antoniadis, Ignatios and Arkani-Hamed, Nima and Dimopoulos, Savas and Dvali, G. R.",
    title = "{New dimensions at a millimeter to a Fermi and superstrings at a TeV}",
    eprint = "hep-ph/9804398",
    archivePrefix = "arXiv",
    reportNumber = "SLAC-PUB-7801, SU-ITP-98-28, CPTH-S608-0498, IC-98-39",
    doi = "10.1016/S0370-2693(98)00860-0",
    journal = "Phys. Lett. B",
    volume = "436",
    pages = "257--263",
    year = "1998"
}

@article{Fang:2018rte,
    author = "Fang, YuHong and Huang, Zhiqi and Miao, HaiTao and Singh, Naveen K.",
    title = {{Anti-evaporation and evaporation of an $n$-dimensional Reissner-Nordstr{\"o}m black hole}},
    eprint = "1811.11547",
    archivePrefix = "arXiv",
    primaryClass = "gr-qc",
    doi = "10.1103/PhysRevD.99.044011",
    journal = "Phys. Rev. D",
    volume = "99",
    number = "4",
    pages = "044011",
    year = "2019"
}

@article{Arkani-Hamed:1998jmv,
    author = "Arkani-Hamed, Nima and Dimopoulos, Savas and Dvali, G. R.",
    title = "{The Hierarchy problem and new dimensions at a millimeter}",
    eprint = "hep-ph/9803315",
    archivePrefix = "arXiv",
    reportNumber = "SLAC-PUB-7769, SU-ITP-98-13",
    doi = "10.1016/S0370-2693(98)00466-3",
    journal = "Phys. Lett. B",
    volume = "429",
    pages = "263--272",
    year = "1998"
}

@article{Reall:2002bh,
    author = "Reall, Harvey S.",
    title = "{Higher dimensional black holes and supersymmetry}",
    eprint = "hep-th/0211290",
    archivePrefix = "arXiv",
    reportNumber = "QMUL-PH-02-20",
    doi = "10.1103/PhysRevD.70.089902",
    journal = "Phys. Rev. D",
    volume = "68",
    pages = "024024",
    year = "2003",
    note = "[Erratum: Phys.Rev.D 70, 089902 (2004)]"
}

@article{Rogatko:2006gg,
    author = "Rogatko, Marek",
    title = "{Classification of Static Charged Black Holes in Higher Dimensions}",
    eprint = "hep-th/0606116",
    archivePrefix = "arXiv",
    doi = "10.1103/PhysRevD.73.124027",
    journal = "Phys. Rev. D",
    volume = "73",
    pages = "124027",
    year = "2006"
}

@article{Myers:1986rx,
    author = "Myers, Robert C.",
    title = "{Higher Dimensional Black Holes in Compactified Space-times}",
    reportNumber = "PRINT-86-0370 (PRINCETON)",
    doi = "10.1103/PhysRevD.35.455",
    journal = "Phys. Rev. D",
    volume = "35",
    pages = "455",
    year = "1987"
}

@article{Elvang:2004rt,
    author = "Elvang, Henriette and Emparan, Roberto and Mateos, David and Reall, Harvey S.",
    title = "{A Supersymmetric black ring}",
    eprint = "hep-th/0407065",
    archivePrefix = "arXiv",
    reportNumber = "NSF-KITP-04-84",
    doi = "10.1103/PhysRevLett.93.211302",
    journal = "Phys. Rev. Lett.",
    volume = "93",
    pages = "211302",
    year = "2004"
}

@article{Tomizawa:2006vp,
    author = "Tomizawa, Shinya and Nozawa, Masato",
    title = "{Vacuum solutions of five-dimensional Einstein equations generated by inverse scattering method. II. Production of black ring solution}",
    eprint = "hep-th/0604067",
    archivePrefix = "arXiv",
    doi = "10.1103/PhysRevD.73.124034",
    journal = "Phys. Rev. D",
    volume = "73",
    pages = "124034",
    year = "2006"
}

@article{Kleihaus:2007kc,
    author = "Kleihaus, Burkhard and Kunz, Jutta and Navarro-Lerida, Francisco",
    editor = {Macias, Alfredo and L{\"a}mmerzahl, Claus and Camacho, Abel},
    title = "{Rotating Black Holes in Higher Dimensions}",
    eprint = "0710.2291",
    archivePrefix = "arXiv",
    primaryClass = "hep-th",
    doi = "10.1063/1.2902801",
    journal = "AIP Conf. Proc.",
    volume = "977",
    number = "1",
    pages = "94--115",
    year = "2008"
}

@article{Emparan:2006mm,
    author = "Emparan, Roberto and Reall, Harvey S.",
    title = "{Black Rings}",
    eprint = "hep-th/0608012",
    archivePrefix = "arXiv",
    doi = "10.1088/0264-9381/23/20/R01",
    journal = "Class. Quant. Grav.",
    volume = "23",
    pages = "R169",
    year = "2006"
}

@article{Galloway:2005mf,
    author = "Galloway, Gregory J. and Schoen, Richard",
    title = "{A Generalization of Hawking's black hole topology theorem to higher dimensions}",
    eprint = "gr-qc/0509107",
    archivePrefix = "arXiv",
    doi = "10.1007/s00220-006-0019-z",
    journal = "Commun. Math. Phys.",
    volume = "266",
    pages = "571--576",
    year = "2006"
}

@article{Myers:1986un,
    author = "Myers, Robert C. and Perry, M. J.",
    title = "{Black Holes in Higher Dimensional Space-Times}",
    reportNumber = "PRINT-86-0067 (PRINCETON)",
    doi = "10.1016/0003-4916(86)90186-7",
    journal = "Annals Phys.",
    volume = "172",
    pages = "304",
    year = "1986"
}

@article{Dymnikova:2010zz,
    author = "Dymnikova, Irina and Korpusik, Michal",
    title = "{Regular black hole remnants in de Sitter space}",
    doi = "10.1016/j.physletb.2010.01.044",
    journal = "Phys. Lett. B",
    volume = "685",
    pages = "12--18",
    year = "2010"
}

@article{Dymnikova:2004zc,
    author = "Dymnikova, Irina",
    title = "{Regular electrically charged structures in nonlinear electrodynamics coupled to general relativity}",
    eprint = "gr-qc/0407072",
    archivePrefix = "arXiv",
    doi = "10.1088/0264-9381/21/18/009",
    journal = "Class. Quant. Grav.",
    volume = "21",
    pages = "4417--4429",
    year = "2004"
}

@article{Burinskii:2002pz,
    author = "Burinskii, Alexander and Hildebrandt, Sergi R.",
    title = "{New type of regular black holes and particle - like solutions from NED}",
    eprint = "hep-th/0202066",
    archivePrefix = "arXiv",
    doi = "10.1103/PhysRevD.65.104017",
    journal = "Phys. Rev. D",
    volume = "65",
    pages = "104017",
    year = "2002"
}

@article{Toshmatov:2019gxg,
    author = "Toshmatov, Bobir and Stuchl{\'\i}k, Zden{\v{e}}k and Ahmedov, Bobomurat and Malafarina, Daniele",
    title = "{Relaxations of perturbations of spacetimes in general relativity coupled to nonlinear electrodynamics}",
    eprint = "1903.03778",
    archivePrefix = "arXiv",
    primaryClass = "gr-qc",
    doi = "10.1103/PhysRevD.99.064043",
    journal = "Phys. Rev. D",
    volume = "99",
    number = "6",
    pages = "064043",
    year = "2019"
}

@article{Junior:2023qaq,
    author = "Junior, Jos{\'e} Tarciso S. S. and Rodrigues, Manuel E.",
    title = "{Coincident $f(\mathbb {Q})$ gravity: black holes, regular black holes, and black bounces}",
    eprint = "2306.04661",
    archivePrefix = "arXiv",
    primaryClass = "gr-qc",
    doi = "10.1140/epjc/s10052-023-11660-2",
    journal = "Eur. Phys. J. C",
    volume = "83",
    number = "6",
    pages = "475",
    year = "2023"
}

@article{Balart:2014cga,
    author = "Balart, Leonardo and Vagenas, Elias C.",
    title = "{Regular black holes with a nonlinear electrodynamics source}",
    eprint = "1408.0306",
    archivePrefix = "arXiv",
    primaryClass = "gr-qc",
    doi = "10.1103/PhysRevD.90.124045",
    journal = "Phys. Rev. D",
    volume = "90",
    number = "12",
    pages = "124045",
    year = "2014"
}

@article{Balart:2014jia,
    author = "Balart, Leonardo and Vagenas, Elias C.",
    title = "{Regular black hole metrics and the weak energy condition}",
    eprint = "1401.2136",
    archivePrefix = "arXiv",
    primaryClass = "gr-qc",
    doi = "10.1016/j.physletb.2014.01.024",
    journal = "Phys. Lett. B",
    volume = "730",
    pages = "14--17",
    year = "2014"
}

@article{Hollenstein:2008hp,
    author = "Hollenstein, Lukas and Lobo, Francisco S. N.",
    title = "{Exact solutions of f(R) gravity coupled to nonlinear electrodynamics}",
    eprint = "0807.2325",
    archivePrefix = "arXiv",
    primaryClass = "gr-qc",
    doi = "10.1103/PhysRevD.78.124007",
    journal = "Phys. Rev. D",
    volume = "78",
    pages = "124007",
    year = "2008"
}

@article{Junior:2023ixh,
    author = "Junior, Jos{\'e} Tarciso S. S. and Lobo, Francisco S. N. and Rodrigues, Manuel E.",
    title = "{(Regular) Black holes in conformal Killing gravity coupled to nonlinear electrodynamics and scalar fields}",
    eprint = "2310.19508",
    archivePrefix = "arXiv",
    primaryClass = "gr-qc",
    doi = "10.1088/1361-6382/ad210e",
    journal = "Class. Quant. Grav.",
    volume = "41",
    number = "5",
    pages = "055012",
    year = "2024"
}

@article{Ahmedov:2021ohg,
    author = "Ahmedov, Bobomurat and Rahimov, Ozodbek and Toshmatov, Bobir",
    title = "{Gravitational Capture Cross-Section of Particles by Schwarzschild-Tangherlini Black Holes}",
    doi = "10.3390/universe7080307",
    journal = "Universe",
    volume = "7",
    number = "8",
    pages = "307",
    year = "2021"
}

@article{Rahimov:2024hol,
    author = "Rahimov, Ozodbek and Toshmatov, Bobir and Vyblyi, Yuriy and Akhmedov, Abdimirkhakim and Abdulazizov, Bahromjon",
    title = {{Charged particle dynamics in Reissner{\textendash}Nordstr{\"o}m{\textendash}Tangherlini spacetime}},
    doi = "10.1016/j.dark.2024.101483",
    journal = "Phys. Dark Univ.",
    volume = "44",
    pages = "101483",
    year = "2024"
}

@article{Dolan:2024qqr,
    author = "Dolan, Sam R. and de Paula, Marco A. A. and Leite, Luiz C. S. and Crispino, Lu{\'\i}s C. B.",
    title = "{Superradiant instability of a charged regular black hole}",
    eprint = "2401.14967",
    archivePrefix = "arXiv",
    primaryClass = "gr-qc",
    doi = "10.1103/PhysRevD.109.124037",
    journal = "Phys. Rev. D",
    volume = "109",
    number = "12",
    pages = "124037",
    year = "2024"
}

@article{dePaula:2024yzy,
    author = "de Paula, Marco A. A. and Lima, Junior., Haroldo C. D. and Cunha, Pedro V. P. and Herdeiro, Carlos A. R. and Crispino, Lu{\'\i}s C. B.",
    title = "{Good tachyons, bad bradyons: Role reversal in Einstein-nonlinear-electrodynamics models}",
    eprint = "2412.18659",
    archivePrefix = "arXiv",
    primaryClass = "gr-qc",
    doi = "10.1016/j.physletb.2025.139513",
    journal = "Phys. Lett. B",
    volume = "866",
    pages = "139513",
    year = "2025"
}

@article{Verbin:2024ewl,
    author = {Verbin, Yosef and Pulice, Beyhan and {\"O}vg{\"u}n, Ali and Huang, Hyat},
    title = "{New black hole solutions of second and first order formulations of nonlinear electrodynamics}",
    eprint = "2412.20989",
    archivePrefix = "arXiv",
    primaryClass = "gr-qc",
    doi = "10.1103/PhysRevD.111.084061",
    journal = "Phys. Rev. D",
    volume = "111",
    number = "8",
    pages = "084061",
    year = "2025"
}

@article{Tlemissov:2025nnk,
    author = "Tlemissov, Abylaikhan and Toshmatov, Bobir and Kov{\'a}{\v{r}}, Ji{\v{r}}{\'\i}",
    title = "{Effect of nonlinear electrodynamics on polarization distribution around black holes}",
    eprint = "2503.08294",
    archivePrefix = "arXiv",
    primaryClass = "gr-qc",
    doi = "10.1103/PhysRevD.111.064084",
    journal = "Phys. Rev. D",
    volume = "111",
    number = "6",
    pages = "064084",
    year = "2025"
}

@article{Huang:2025uhv,
    author = "Huang, Hyat and Rao, Xiao-Ping",
    title = "{Regular black holes and their singular families}",
    eprint = "2503.13133",
    archivePrefix = "arXiv",
    primaryClass = "gr-qc",
    doi = "10.1103/PhysRevD.111.104040",
    journal = "Phys. Rev. D",
    volume = "111",
    number = "10",
    pages = "104040",
    year = "2025"
}

@article{Capozziello:2025ycu,
    author = "Capozziello, Salvatore and Gambino, Serena and Luongo, Orlando",
    title = "{Comparing Bondi and Novikov{\textendash}Thorne accretion disk luminosity around regular black holes}",
    eprint = "2503.21987",
    archivePrefix = "arXiv",
    primaryClass = "gr-qc",
    doi = "10.1016/j.dark.2025.101950",
    journal = "Phys. Dark Univ.",
    volume = "48",
    pages = "101950",
    year = "2025"
}

@article{Suvorov:2025lar,
    author = "Suvorov, Arthur G. and Bargue{\~n}o, Pedro",
    title = "{Doubly regular black holes}",
    eprint = "2507.23250",
    archivePrefix = "arXiv",
    primaryClass = "gr-qc",
    doi = "10.1103/ljjy-v4m1",
    journal = "Phys. Rev. D",
    volume = "112",
    number = "4",
    pages = "044027",
    year = "2025"
}

@article{Chen:2025aom,
    author = "Chen, Che-Yu and De Felice, Antonio and Tsujikawa, Shinji and Sano, Taishi",
    title = "{Vector Horndeski black holes in nonlinear electrodynamics}",
    eprint = "2509.23134",
    archivePrefix = "arXiv",
    primaryClass = "gr-qc",
    reportNumber = "RIKEN-iTHEMS-Report-25, YITP-25-152, WUCG-25-11",
    doi = "10.1103/fjqh-7gb2",
    journal = "Phys. Rev. D",
    volume = "113",
    number = "2",
    pages = "024027",
    year = "2026"
}

@article{Bokulic:2025brf,
    author = "Bokuli{\'c}, Ana and Juri{\'c}, Tajron and Smoli{\'c}, Ivica",
    title = "{Conundrum of regular black holes with nonlinear electromagnetic fields}",
    eprint = "2510.23711",
    archivePrefix = "arXiv",
    primaryClass = "gr-qc",
    reportNumber = "ZTF-EP-25-07; RBI-ThPhys-2025-41",
    doi = "10.1103/z7gd-96ms",
    journal = "Phys. Rev. D",
    volume = "113",
    number = "2",
    pages = "024044",
    year = "2026"
}

@article{Bueno:2024zsx,
    author = "Bueno, Pablo and Cano, Pablo A. and Hennigar, Robie A. and Murcia, {\'A}ngel J.",
    title = "{Regular black holes from thin-shell collapse}",
    eprint = "2412.02740",
    archivePrefix = "arXiv",
    primaryClass = "gr-qc",
    doi = "10.1103/PhysRevD.111.104009",
    journal = "Phys. Rev. D",
    volume = "111",
    number = "10",
    pages = "104009",
    year = "2025"
}

@article{Guerrero:2020uhn,
    author = "Guerrero, Merce and Rubiera-Garcia, Diego",
    title = "{Nonsingular black holes in nonlinear gravity coupled to Euler-Heisenberg electrodynamics}",
    eprint = "2005.08828",
    archivePrefix = "arXiv",
    primaryClass = "gr-qc",
    doi = "10.1103/PhysRevD.102.024005",
    journal = "Phys. Rev. D",
    volume = "102",
    number = "2",
    pages = "024005",
    year = "2020"
}

@article{Aros:2019quj,
    author = "Aros, Rodrigo and Estrada, Milko",
    title = "{Regular black holes and its thermodynamics in Lovelock gravity}",
    eprint = "1901.08724",
    archivePrefix = "arXiv",
    primaryClass = "gr-qc",
    doi = "10.1140/epjc/s10052-019-6783-7",
    journal = "Eur. Phys. J. C",
    volume = "79",
    number = "3",
    pages = "259",
    year = "2019"
}

@article{Chamseddine:2016ktu,
    author = "Chamseddine, Ali H. and Mukhanov, Viatcheslav",
    title = "{Nonsingular Black Hole}",
    eprint = "1612.05861",
    archivePrefix = "arXiv",
    primaryClass = "gr-qc",
    doi = "10.1140/epjc/s10052-017-4759-z",
    journal = "Eur. Phys. J. C",
    volume = "77",
    number = "3",
    pages = "183",
    year = "2017"
}

@article{Balakin:2015gpq,
    author = "Balakin, Alexander B. and Lemos, Jos{\'e} P. S. and Zayats, Alexei E.",
    title = "{Regular nonminimal magnetic black holes in spacetimes with a cosmological constant}",
    eprint = "1512.02653",
    archivePrefix = "arXiv",
    primaryClass = "gr-qc",
    doi = "10.1103/PhysRevD.93.024008",
    journal = "Phys. Rev. D",
    volume = "93",
    number = "2",
    pages = "024008",
    year = "2016"
}

@article{Nicolini:2005vd,
    author = "Nicolini, Piero and Smailagic, Anais and Spallucci, Euro",
    title = "{Noncommutative geometry inspired Schwarzschild black hole}",
    eprint = "gr-qc/0510112",
    archivePrefix = "arXiv",
    doi = "10.1016/j.physletb.2005.11.004",
    journal = "Phys. Lett. B",
    volume = "632",
    pages = "547--551",
    year = "2006"
}

@article{Lemos:2011dq,
    author = "Lemos, Jose P. S. and Zanchin, Vilson T.",
    title = {{Regular black holes: Electrically charged solutions, Reissner-Nordstr{\"o}m outside a de Sitter core}},
    eprint = "1104.4790",
    archivePrefix = "arXiv",
    primaryClass = "gr-qc",
    doi = "10.1103/PhysRevD.83.124005",
    journal = "Phys. Rev. D",
    volume = "83",
    pages = "124005",
    year = "2011"
}

@article{Bueno:2024dgm,
    author = "Bueno, Pablo and Cano, Pablo A. and Hennigar, Robie A.",
    title = "{Regular black holes from pure gravity}",
    eprint = "2403.04827",
    archivePrefix = "arXiv",
    primaryClass = "gr-qc",
    doi = "10.1016/j.physletb.2025.139260",
    journal = "Phys. Lett. B",
    volume = "861",
    pages = "139260",
    year = "2025"
}

@article{Bronnikov:PRL:2000,
author = {Bronnikov, K.},
year = {2000},
month = {12},
pages = {4641},
title = "{Comment on "Regular Black Hole in General Relativity Coupled to Nonlinear Electrodynamics"}",
volume = {85},
journal = {Phys. Rev. Lett.},
doi = {10.1103/PhysRevLett.85.4641}
}

@ARTICLE{Bronnikov:PRD:2001,
   author = {{Bronnikov}, K.~A.},
    title = "{Regular magnetic black holes and monopoles from nonlinear electrodynamics}",
  journal = {Phys. Rev. D},
   eprint = {gr-qc/0006014},
     year = 2001,
    month = feb,
   volume = 63,
   number = 4,
      eid = {044005},
    pages = {044005},
      doi = {10.1103/PhysRevD.63.044005},
   adsurl = {http://adsabs.harvard.edu/abs/2001PhRvD..63d4005B}
}

@article{Toshmatov:2021fgm,
    author = "Toshmatov, Bobir and Ahmedov, Bobomurat and Malafarina, Daniele",
    title = "{Can a light ray distinguish charge of a black hole in nonlinear electrodynamics?}",
    eprint = "2101.05496",
    archivePrefix = "arXiv",
    primaryClass = "gr-qc",
    doi = "10.1103/PhysRevD.103.024026",
    journal = "Phys. Rev. D",
    volume = "103",
    number = "2",
    pages = "024026",
    year = "2021"
}

@article{Salazar:1987ap,
    author = "Salazar, I. H. and Garcia, A. and Plebanski, J.",
    title = "{Duality Rotations and Type $D$ Solutions to Einstein Equations With Nonlinear Electromagnetic Sources}",
    doi = "10.1063/1.527430",
    journal = "J. Math. Phys.",
    volume = "28",
    pages = "2171--2181",
    year = "1987"
}

@misc{rasheed2003,
      title={Non-Linear Electrodynamics: Zeroth and First Laws of Black Hole Mechanics}, 
      author={D. A. Rasheed},
      year={2003},
      eprint={hep-th/9702087},
      archivePrefix={arXiv},
      primaryClass={hep-th},
      url={https://arxiv.org/abs/hep-th/9702087}, 
}

@article{Breton:2004qa,
    author = "Breton, Nora",
    title = "{Smarr's formula for black holes with non-linear electrodynamics}",
    eprint = "gr-qc/0405116",
    archivePrefix = "arXiv",
    doi = "10.1007/s10714-005-0051-x",
    journal = "Gen. Rel. Grav.",
    volume = "37",
    pages = "643--650",
    year = "2005"
}

@article{Zhang:2016ilt,
    author = "Zhang, Yuan and Gao, Sijie",
    title = "{First law and Smarr formula of black hole mechanics in nonlinear gauge theories}",
    eprint = "1610.01237",
    archivePrefix = "arXiv",
    primaryClass = "gr-qc",
    doi = "10.1088/1361-6382/aac9d4",
    journal = "Class. Quant. Grav.",
    volume = "35",
    number = "14",
    pages = "145007",
    year = "2018"
}

@article{Ma:2014qma,
    author = "Ma, Meng-Sen and Zhao, Ren",
    title = "{Corrected form of the first law of thermodynamics for regular black holes}",
    eprint = "1411.0833",
    archivePrefix = "arXiv",
    primaryClass = "gr-qc",
    doi = "10.1088/0264-9381/31/24/245014",
    journal = "Class. Quant. Grav.",
    volume = "31",
    pages = "245014",
    year = "2014"
}

@article{Malafarina:2022oka,
    author = "Malafarina, Daniele and Toshmatov, Bobir",
    title = "{Connection between regular black holes in nonlinear electrodynamics and semiclassical dust collapse}",
    eprint = "2204.04025",
    archivePrefix = "arXiv",
    primaryClass = "gr-qc",
    doi = "10.1103/PhysRevD.105.L121502",
    journal = "Phys. Rev. D",
    volume = "105",
    number = "12",
    pages = "L121502",
    year = "2022"
}

@article{Bonanno:2023rzk,
    author = "Bonanno, Alfio and Malafarina, Daniele and Panassiti, Antonio",
    title = "{Dust Collapse in Asymptotic Safety: A Path to Regular Black Holes}",
    eprint = "2308.10890",
    archivePrefix = "arXiv",
    primaryClass = "gr-qc",
    doi = "10.1103/PhysRevLett.132.031401",
    journal = "Phys. Rev. Lett.",
    volume = "132",
    number = "3",
    pages = "031401",
    year = "2024"
}

@article{Misner:1964je,
    author = "Misner, Charles W. and Sharp, David H.",
    title = "{Relativistic equations for adiabatic, spherically symmetric gravitational collapse}",
    doi = "10.1103/PhysRev.136.B571",
    journal = "Phys. Rev.",
    volume = "136",
    pages = "B571--B576",
    year = "1964"
}

@article{Cai:2009qf,
    author = "Cai, Rong-Gen and Cao, Li-Ming and Hu, Ya-Peng and Ohta, Nobuyoshi",
    title = "{Generalized Misner-Sharp Energy in f(R) Gravity}",
    eprint = "0910.2387",
    archivePrefix = "arXiv",
    primaryClass = "hep-th",
    reportNumber = "KU-TP-036",
    doi = "10.1103/PhysRevD.80.104016",
    journal = "Phys. Rev. D",
    volume = "80",
    pages = "104016",
    year = "2009"
}

@article{Nielsen:2008kd,
    author = "Nielsen, Alex B. and Yeom, Dong-han",
    title = "{Spherically symmetric trapping horizons, the Misner-Sharp mass and black hole evaporation}",
    eprint = "0804.4435",
    archivePrefix = "arXiv",
    primaryClass = "gr-qc",
    doi = "10.1142/S0217751X09045984",
    journal = "Int. J. Mod. Phys. A",
    volume = "24",
    pages = "5261--5285",
    year = "2009"
}

@article{Panassiti:2025diw,
    author = "Panassiti, Antonio",
    title = "{Regular black hole cores via gravitational evanescence of collapsing matter}",
    eprint = "2509.17234",
    archivePrefix = "arXiv",
    primaryClass = "gr-qc",
    doi = "10.1103/fbz2-8n2h",
    journal = "Phys. Rev. D",
    volume = "113",
    number = "6",
    pages = "064057",
    year = "2026"
}

@article{Maeda:2007uu,
    author = "Maeda, Hideki and Nozawa, Masato",
    title = "{Generalized Misner-Sharp quasi-local mass in Einstein-Gauss-Bonnet gravity}",
    eprint = "0709.1199",
    archivePrefix = "arXiv",
    primaryClass = "hep-th",
    reportNumber = "CECS-PHY-07-10",
    doi = "10.1103/PhysRevD.77.064031",
    journal = "Phys. Rev. D",
    volume = "77",
    pages = "064031",
    year = "2008"
}

@ARTICLE{Toshmatov:PRD:comment,
       author = {{Toshmatov}, Bobir and {Stuchl{\'\i}k}, Zden{\v{e}}k and
         {Ahmedov}, Bobomurat},
        title = "{Comment on ``Construction of regular black holes in general relativity''}",
      journal = {Phys. Rev. D},
         year = 2018,
        month = jul,
       volume = {98},
       number = {2},
          eid = {028501},
        pages = {028501},
          doi = {10.1103/PhysRevD.98.028501},
archivePrefix = {arXiv},
       eprint = {1807.09502},
 primaryClass = {gr-qc},
       adsurl = {https://ui.adsabs.harvard.edu/abs/2018PhRvD..98b8501T}
}

@ARTICLE{Fan:PRD:2016,
       author = {{Fan}, Zhong-Ying and {Wang}, Xiaobao},
        title = "{Construction of regular black holes in general relativity}",
      journal = {Phys. Rev. D},
         year = 2016,
        month = dec,
       volume = {94},
       number = {12},
          eid = {124027},
        pages = {124027},
          doi = {10.1103/PhysRevD.94.124027},
archivePrefix = {arXiv},
       eprint = {1610.02636},
 primaryClass = {gr-qc},
       adsurl = {https://ui.adsabs.harvard.edu/abs/2016PhRvD..94l4027F}
}

@article{Bronnikov17,
      author         = "Bronnikov, Kirill A.",
      title          = "{Comment on “Construction of regular black holes in
                        general relativity”}",
      journal        = "Phys. Rev. D",
      volume         = "96",
      year           = "2017",
      number         = "12",
      pages          = "128501",
      doi            = "10.1103/PhysRevD 96.128501",
      eprint         = "1712.04342",
      archivePrefix  = "arXiv",
      primaryClass   = "gr-qc",
      SLACcitation   = "%%CITATION = ARXIV:1712.04342;%%"
}

@ARTICLE{Toshmatov17,
   author = {{Toshmatov}, B. and {Stuchl{\'{\i}}k}, Z. and {Ahmedov}, B.},
    title = "{Generic rotating regular black holes in general relativity coupled to nonlinear electrodynamics}",
  journal = {Phys. Rev. D},
archivePrefix = "arXiv",
   eprint = {1704.07300},
 primaryClass = "gr-qc",
     year = 2017,
    month = apr,
   volume = 95,
   number = 8,
      eid = {084037},
    pages = {084037},
      doi = {10.1103/PhysRevD.95.084037},
   adsurl = {http://adsabs.harvard.edu/abs/2017PhRvD..95h4037T}
}

@article{Croney2025,
   title={Thermodynamics of dyonic black holes in non-linear electrodynamics},
   volume={2025},
   ISSN={1029-8479},
   url={http://dx.doi.org/10.1007/JHEP10(2025)013},
   DOI={10.1007/jhep10(2025)013},
   number={10},
   journal={J. High Energy Phys.},
   publisher={Springer Science and Business Media LLC},
   author={Croney, Lewis and Gregory, Ruth and Ramírez-Valdez, Carlos J.},
   year={2025},
   month=Oct }

@article{Bokuli2021,
   title={Black hole thermodynamics in the presence of nonlinear electromagnetic fields},
   volume={103},
   ISSN={2470-0029},
   url={http://dx.doi.org/10.1103/PhysRevD.103.124059},
   DOI={10.1103/physrevd.103.124059},
   number={12},
   journal={Phys. Rev. D},
   publisher={American Physical Society (APS)},
   author={Bokulić, A. and Smolić, I. and Jurić, T.},
   year={2021},
   month=June }

@ARTICLE{Ayon-Beato98,
   author = {{Ay{\'o}n-Beato}, E. and {Garc{\'{\i}}a}, A.},
    title = "{Regular Black Hole in General Relativity Coupled to Nonlinear Electrodynamics}",
  journal = {Phys. Rev. Lett.},
   eprint = {gr-qc/9911046},
     year = 1998,
    month = jun,
   volume = 80,
    pages = {5056-5059},
      doi = {10.1103/PhysRevLett.80.5056},
   adsurl = {http://adsabs.harvard.edu/abs/1998PhRvL..80.5056A}
}

@ARTICLE{Ayon-Beato00,
   author = {{Ay{\'o}n-Beato}, E. and {Garc{\'{\i}}a}, A.},
    title = "{The Bardeen model as a nonlinear magnetic monopole}",
  journal = {Phys. Lett. B},
   eprint = {gr-qc/0009077},
     year = 2000,
    month = nov,
   volume = 493,
    pages = {149-152},
      doi = {10.1016/S0370-2693(00)01125-4},
   adsurl = {http://adsabs.harvard.edu/abs/2000PhLB..493..149A}
}

@article{Gulin:2017ycu,
    author = "Gulin, Luka and Smoli{\'c}, Ivica",
    title = "{Generalizations of the Smarr formula for black holes with nonlinear electromagnetic fields}",
    journal = "Class. Quant. Grav.",
    volume = "35",
    number = "2",
    pages = "025015",
    year = "2018",
    eprint = "1710.04660",
    archivePrefix = "arXiv",
    primaryClass = "gr-qc",
    doi = "10.1088/1361-6382/aa9dfd"
}

@ARTICLE{letter,
       author = {{Toshmatov}, Bobir and {Ahmedov}, Bobomurat and {Isamadinova}, Nozima and {Yuan}, Chengxun},
        title = "{Regular black holes do not violate the first law of thermodynamics}",
      journal = {arXiv e-prints},
         year = 2026,
        month = oct,
          eid = {arXiv:2610.08099},
        pages = {arXiv:2610.08099},
          doi = {10.48550/arXiv.2610.08099},
archivePrefix = {arXiv},
       eprint = {2610.08099},
 primaryClass = {gr-qc},
       adsurl = {https://ui.adsabs.harvard.edu/abs/2026arXiv261008099T}
}

\end{document}